\documentclass[journal]{IEEEtran}
\usepackage{graphicx}
\usepackage{amsmath}
\usepackage{multirow}
\usepackage{colortbl}
\usepackage{color}
\usepackage{soul}
\usepackage{enumerate}
\usepackage[caption=false]{subfig}
\usepackage[table]{xcolor}
\usepackage{pifont}
\usepackage{paralist}
\usepackage{siunitx}
\usepackage{siunitx}
\usepackage{flushend}
\usepackage{amssymb}
\usepackage{paralist}
\usepackage{siunitx}
\usepackage{epstopdf}
\usepackage[inline]{enumitem}
\IEEEoverridecommandlockouts

\usepackage{textcomp}
\usepackage{stfloats}
\usepackage{tikz}
\usepackage{verbatim}
\usetikzlibrary{chains,positioning,shapes.symbols,arrows,shapes,shadows,trees}
\usepackage{standalone}
\usepackage{array}
\usepackage{cite}
\usepackage{cuted, nccmath}
\usepackage{lipsum}
\usepackage{cleveref}

\begin{document}
\bstctlcite{IEEEexample:BSTcontrol}
\setlength{\parskip}{0pt}

\title{Graphene and Related Materials\\ for the Internet of Bio-Nano Things}

\author{
Meltem Civas$^\ast$, Murat Kuscu$^\ast$, Oktay Cetinkaya, Beyza E. Ortlek, Ozgur B. Akan
      \thanks{The authors are with the Department of Electrical and Electronics Engineering, Ko\c{c} University, Istanbul, 34450, Turkey (email: \{mcivas16, mkuscu, ocetinkaya, bortlek14, sfathipour21, akan\}@ku.edu.tr).}
       \thanks{Ozgur B. Akan is also with the Internet of Everything (IoE) Group, Electrical Engineering Division, Department of Engineering, University of Cambridge, Cambridge, CB3 0FA, UK (email: oba21@cam.ac.uk).}
       
\thanks{This work was supported in part by the AXA Research Fund (AXA Chair for Internet of Everything at Ko\c{c} University), The Scientific and Technological Research Council of Turkey (TUBITAK) under Grant \#120E301, European
Union’s Horizon 2020 Research and Innovation Programme through the Marie Skłodowska-Curie Individual Fellowship under Grant Agreement \#101028935, and Huawei Graduate Research Scholarship.}
\thanks{$^\ast$These authors equally contributed to this work.}

}

\maketitle

\begin{abstract}
Internet of Bio-Nano Things (IoBNT) is a transformative communication framework, characterized by heterogeneous networks comprising both biological entities and artificial micro/nano-scale devices, so-called Bio-Nano Things (BNTs), interfaced with conventional communication networks for enabling innovative biomedical and environmental applications. Realizing the potential of IoBNT requires the development of new and unconventional communication technologies, such as molecular communications, as well as the corresponding transceivers, bio-cyber interfacing technologies connecting the biochemical domain of IoBNT to the electromagnetic domain of conventional networks, and miniaturized energy harvesting and storage components for the continuous power supply to BNTs. Graphene and related materials (GRMs) exhibit exceptional electrical, optical, biochemical, and mechanical properties, rendering them ideal candidates for addressing the challenges posed by IoBNT. This perspective article highlights recent advancements in GRM-based device technologies that are promising for implementing the core components of IoBNT. By identifying the unique opportunities afforded by GRMs and aligning them with the practical challenges associated with IoBNT, particularly in the materials domain, our aim is to accelerate the transition of envisaged IoBNT applications from theoretical concepts to practical implementations, while also uncovering new application areas for GRMs.
\end{abstract}


\section{Introduction}
The contemporary landscape of information and communication technologies (ICT) is experiencing a rapid evolution, driven by the proliferation of Internet of Things (IoT) applications. These applications hinge on the interconnection of a vast number of heterogeneous devices that interact closely with the physical world, utilizing an array of sensing and communication technologies. Advancements in biotechnology and nanotechnology, particularly in synthetic biology and nanomaterials, pave the way for the next significant milestone: extending our control and connectivity to the universal scale, including the complex biotic world, ultimately serving the holistic \emph{Internet of Everything (IoE)} vision \cite{civas2021universal, dinc2019internet, akan2022internet}. 

\begin{figure*}[t]
	\centering
	\includegraphics[width=1\textwidth]{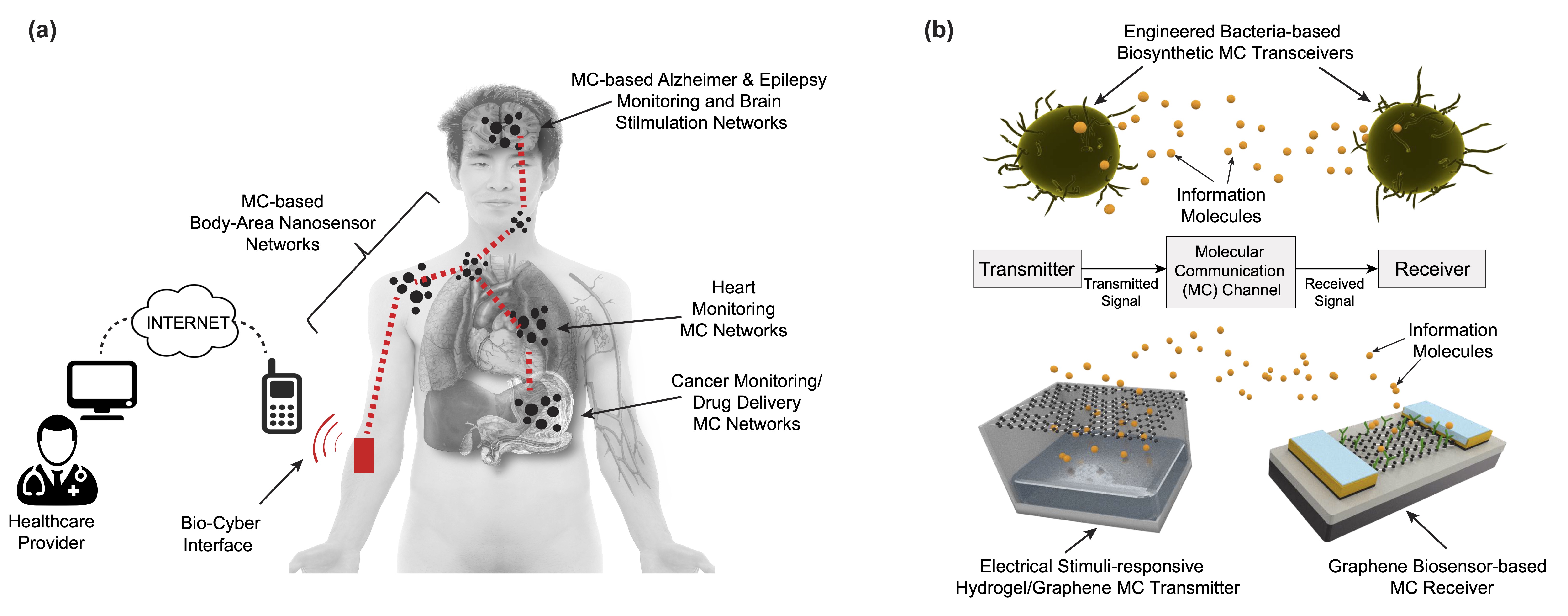}
	\caption{Conceptual drawing of a continuous health monitoring application of IoBNT. (b) Molecular communications among engineered bacteria, and between GRM-based BNTs \cite{kuscu2021internet}.}
	\label{fig:IoBNT}
\end{figure*}

Central to the IoE vision is the emerging \emph{Internet of Bio-Nano Things (IoBNT)} framework, which encompasses collaborative networks of \emph{Bio-Nano Things (BNTs)}. These networks consist of diverse array of context-aware entities, including biological beings and artificial micro/nanoscale devices, such as engineered bacteria and nanosensors that interact closely with their natural counterparts in complex biological environments \cite{akyildiz2015internet}. The IoBNT framework also envisions integrating BNTs into conventional communication networks through seamless bio-cyber interfaces, thereby enabling novel applications, such as remote and continuous health monitoring with nanobiosensor networks inside human body (Fig. \ref{fig:IoBNT}(a)), theranostic smart drug delivery systems that can be remotely monitored and controlled, distributed sensing of chemical agents with engineered bacteria networks in various environments, artificial biochemical networks that replace or reinforce their natural counterparts for treatment of diseases, and many others \cite{khan2020nanosensor, akan2021information}.  

The biotic world is characterized by immense heterogeneity among interacting entities, such as living cells, animals, plants, and bacterial communities, communicating across various spatio-temporal scales in challenging biological environments, such as inside human body. These natural BNTs interact through unconventional communication methods, such as exchanging biochemical and electrochemical signals, across diverse physiological environments \cite{kuscu2021internet}. 
At the fundamental level, biological interactions primarily involve micro/nanoscale components or their modular assemblies, such as ligand receptors and cargo-carrying microtubules \cite{mann2008life}, highlighting the essential role of artificial BNTs within the IoBNT framework for monitoring and controlling biological systems at single molecular resolution. 

IoBNT presents numerous challenges, foremost of which is the need for innovative communication technologies that can withstand noisy and dynamic media such as biological environments. These technologies must be adaptable to micro/nano size scales and diverse biotic and abiotic characteristics of BNTs. Conventional electromagnetic (EM) communications are considered unsuitable due to size mismatch and BNTs' limited energy and computational resources. Emerging unconventional communication technologies offer potential solutions for enabling heterogeneous networks in IoBNT. Bio-inspired molecular communications (MC) stand out as a prime method for interconnecting BNTs, delivering superior biocompatibility and biotic-abiotic interoparability by harnessing nature's communication strategies. Terahertz (THz)-band EM communications and ultrasonic nanocommunications, utilizing extremely high-frequency EM waves and high-frequency acoustic waves, respectively, represent additional promising IoBNT communication technologies closely aligned with conventional ICT expertise. 

Unconventional communication methods necessitate novel transceiver (Tx/Rx) architectures that are also unconventional in size, complexity, and transmission and reception functionalities. In MC, for instance, transceivers should be capable of encoding information into distinguishable molecular properties, such as molecular concentration, transmitting molecular signals into the channel, and subsequently sensing and decoding these signals \cite{akan2016fundamentals} (Fig. \ref{fig:IoBNT}(b)). The design of transceivers for THz-band EM nanocommunications necessitate revisiting the antenna theory to accommodate plasmons at the interface between solid-state materials and high-frequency airborne EM waves \cite{jornet2014graphene}. Fortunately, recent advancements in nanomaterials and nanofabrication techniques have facilitated significant progress in designing and fabricating practical Tx/Rxs for unconventional communication modalities. In particular, graphene and related materials (GRMs), with their exceptional electrical, optical, biochemical and mechanical properties, biocompatibility, and ease of fabrication and integration, show great potential in this area. Graphene's high surface-to-volume ratio makes it highly sensitive to biomolecules, enabling detection of a wide variety of molecular signals in biological environments. Additionally, graphene's plasmonic response supports the emission and detection of THz radiation \cite{tamagnone2012reconfigurable, bandurin2018resonant}, while its exceptional thermo-photoacoustic properties enable wideband emission of high-frequency ultrasonic signals \cite{hwan2012reduced}.  

Interfacing BNTs and IoBNT networks, which partially or entirely rely on the exchange of molecular signals in challenging biological environments, with conventional EM-based communication networks presents another significant challenge, necessitating bio-cyber or nano-macro interfaces. These interfaces translate biochemical signals into other physical forms, such as electrical, optical, or EM, that can be processed by conventional ICT devices, vice versa.  Existing neural interfaces, electrical biosensors, and drug delivery devices offer valuable insights for designing bio-cyber interfaces for IoBNT. Nanomaterials, particularly GRMs, stand out in these emerging technologies because of their exceptional ability to convert signals between various forms. Inspiring examples include graphene field-effect transistor (FET) sensors transducing biomolecular concentrations into electrical signals \cite{kwong2019chemically}, and GRM nanocomposites releasing drug molecules upon electrical stimulation \cite{weaver2014electrically}. 

Energy self-sufficiency is crucial for most IoBNT applications, requiring BNTs to harvest energy from ambient sources and store it for continuous operation. The development of miniaturized energy harvesters and storage components relies heavily on nanomaterials and micro/nanofabrication technologies, which can also facilitate on-chip integration for IoBNT components. GRMs, with their remarkable electromechanical properties and flexibility, have found widespread use in these technologies. Their exceptional electrical and thermal conductivity, and mechanical strength make them suitable for harnessing various ambient energy sources, including mechanical vibrations, sound, heat, and light. Moreover, GRMs with large specific surface area, high electrical conductivity, and electrochemical stability exhibit great potential as high-performance active electrode materials for micro-supercapacitors (MSCs), an emerging technology for realizing integrable micro/nanoscale energy storage components for BNTs.

Therefore, GRMs with outstanding properties hold great promise in addressing the key challenges of IoBNT. In light of this, this paper evaluates the potential of GRMs and related device technologies for the practical implementation of fundamental IoBNT components, namely micro/nanoscale transceivers, bio-cyber interfaces, and energy harvesting and storage components. By evaluating these technologies, we aim to contribute to the development and realization of the IoBNT vision, thereby fostering the emergence of groundbreaking applications at the intersection of nanotechnology, biotechnology, and ICT. 

\section{Graphene and Related Materials (GRMs)}
\label{GRM}

Graphene, a two-dimensional (2D) monolayer of carbon atoms arranged in a honeycomb crystal lattice, has captivated researchers due to its extraordinary electrical, optical, thermal, mechanical, and biochemical properties. These properties make graphene an excellent alternative for existing materials in a wide range of nanotechnology applications. The first successful isolation of graphene from graphite and the subsequent groundbreaking experiments earned the researchers a Nobel Prize in 2010 \cite{novoselov_electric_2004}. In this section, we provide an overview of the fundamental properties of GRMs relevant for the fabrication of abiotic BNTs and IoBNT applications, as well as a discussion of widely-used GRM synthesis methods.

\subsection{Properties of GRMs}
\begin{figure}[t]
	\centering
	\includegraphics[width=1\columnwidth]{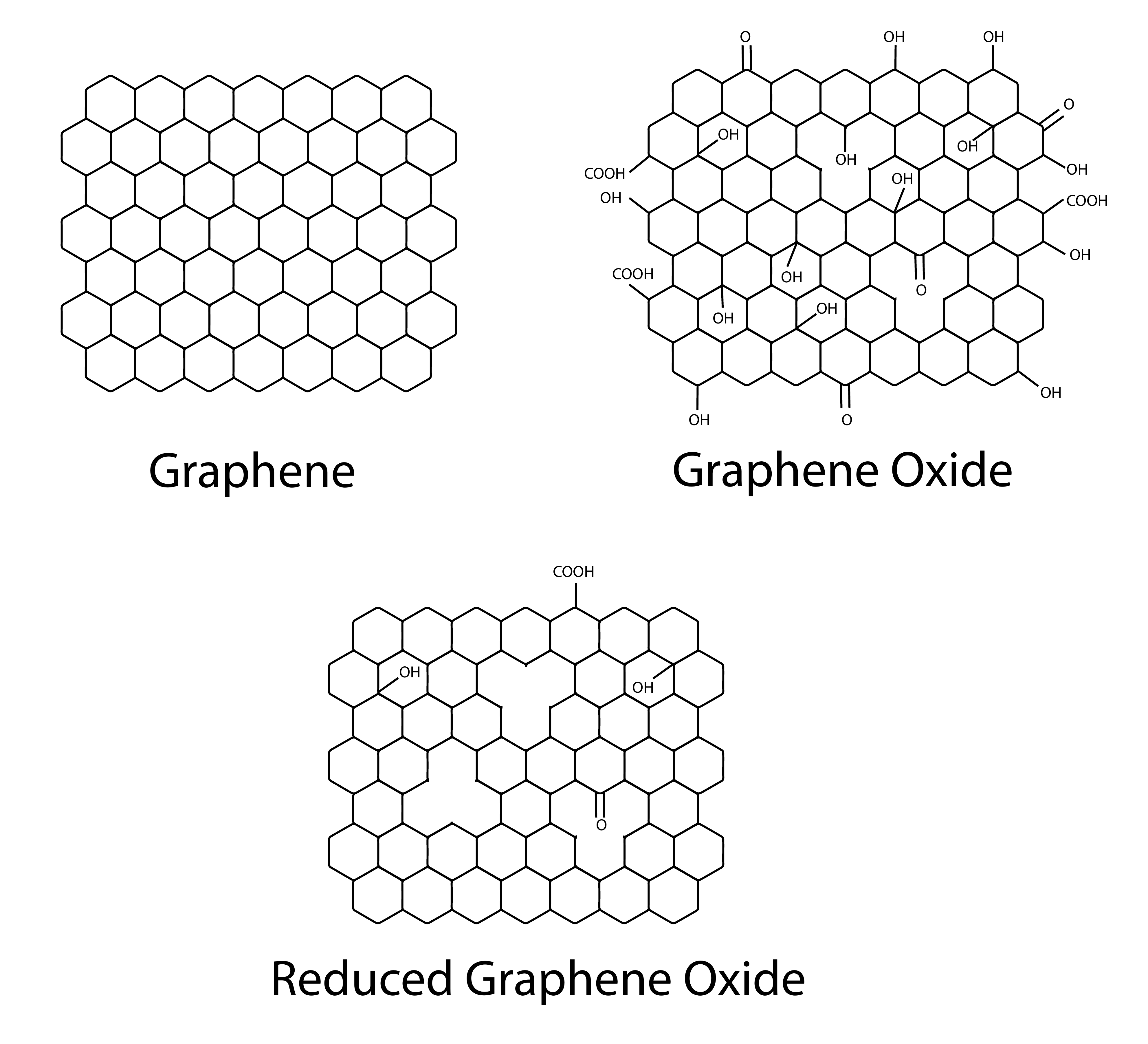}
	\caption{Chemical structures of graphene, graphene oxide, and reduced graphene oxide.}
	\label{fig:GRMs}
\end{figure}

Graphene is the very first example of a closely-packed 2D material isolated in nature. The $sp^2$ hybridization of carbon (C) atoms in graphene results in a unique honeycomb structure (Fig. \ref{fig:GRMs}), where each C atom connects to three other C atoms at a 120-degree angle and with an inter-atomic distance of 1.42 \r{A}. This structure leaves one electron in the out-of-plane $\pi$ orbital, which can move almost freely in the lattice due to the large overlap between the $\pi$ orbitals of neighboring C atoms \cite{zhen_structure_2018}. The strong in-plane $\sigma_{C-C}$ bonds give rise to graphene's remarkable mechanical strength, while the out-of-plane $\pi$ orbitals account for its extraordinary carrier mobility \cite{liao_chemistry_2014}. Graphene’s unique structure serves as the basis for other graphitic carbon allotropes, including graphite, carbon nanotubes (CNTs), fullerene, graphyne, and related materials (e.g., charcoal, carbon fiber (CF), amorphous carbon (AC)). 

Graphene is the mother of GRMs, which primarily consist of single-layer graphene (SLG), bilayer graphene (BLG), few-layer graphene (FLG), graphene oxide (GO), reduced graphene oxide (rGO), graphene nanoribbons (GNRs), graphene nanoplatelets (GNPs), and others \cite{kovtun2019benchmarking}. BLG and FLG consist of two and more than two layers of hexagonal lattices, respectively, held together by weak van der Waals forces. GO shares a similar honeycomb structure with graphene (Fig. \ref{fig:GRMs}), but combines $sp^2$ and $sp^3$ hybridizations. GO can be produced through chemical oxidation of graphite, which results in functionalization with oxygen-containing functional groups \cite{li_graphene-based_2008}. Reduction of GO, either chemically or through thermal annealing, yields rGO, which resembles pristine graphene but with more defects and remnants of oxygen-groups on the surface (Fig. \ref{fig:GRMs}). Furthermore, GNRs are planar quasi-one-dimensional (1D) graphene structures, which can be further categorized into two groups based on edge termination: 1) armchair and 2) zigzag. While zigzag GNRs exhibit metallic properties, armchair GNRs can be metallic or semiconducting, depending on their width \cite{afshari_electronic_2017}.

\begin{figure}[t]
	\centering
	\includegraphics[width=0.8\columnwidth]{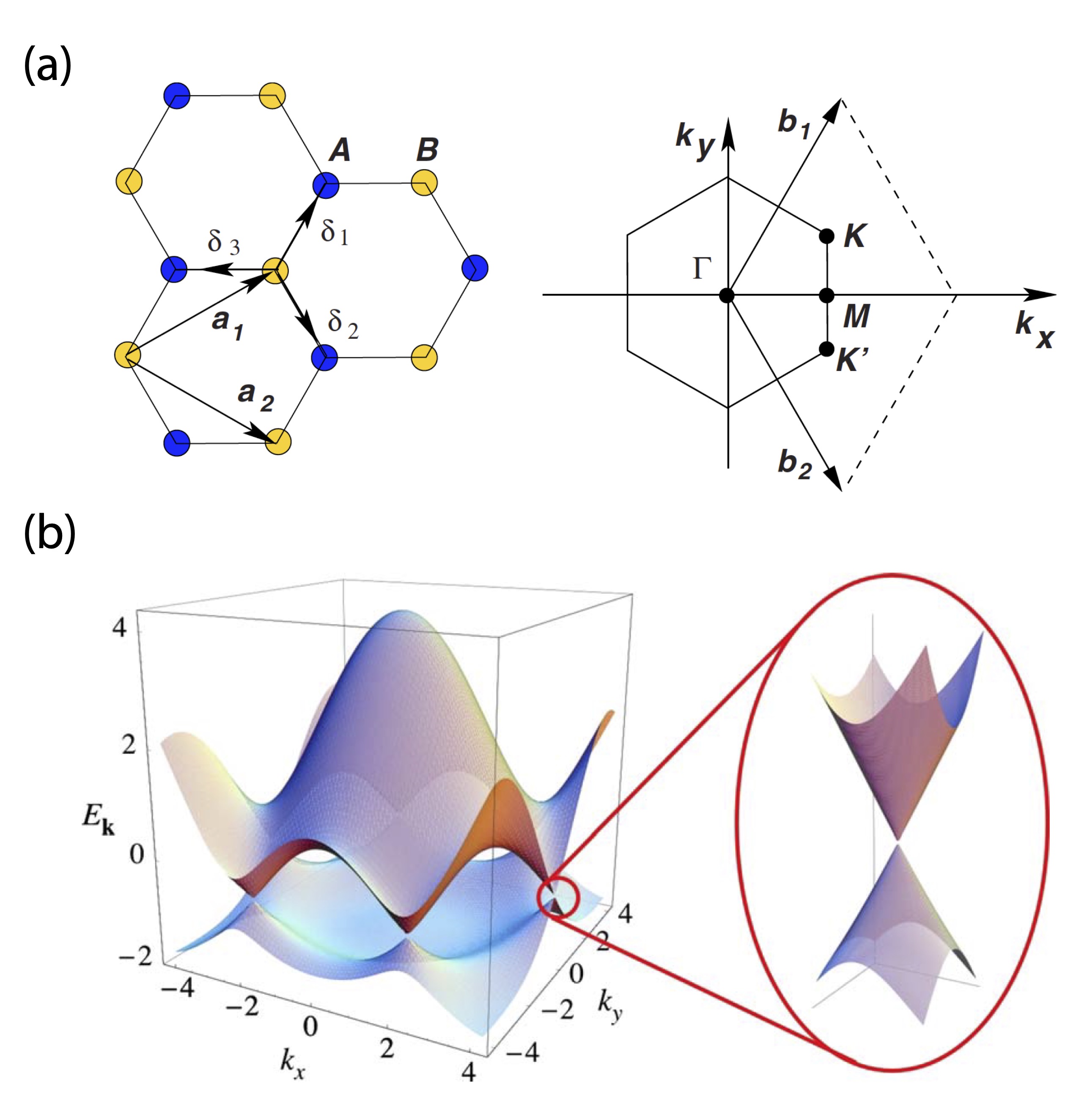}
	\caption{(a) Graphene's honeycomb lattice (left) and the corresponding Brillouin zone in the reciprocal space (right). (b) Energy-momentum dispersion of graphene with a closer look at energy bands near Dirac point. Reproduced with permission from \cite{neto2009electronic}.}
	\label{fig:grapheneElectronicStructure}
\end{figure}

The hexagonal lattice structure of graphene, with two identical atoms in its unit cell, results in a distinct energy-momentum dispersion relation characterized by a conical energy band diagram near the Brillouin zone corners, where the valence and conduction bands touch each other, as depicted in Fig. \ref{fig:grapheneElectronicStructure}. This electronic band structure, with a zero band gap, sets graphene apart from conventional semiconductors and metals, placing it in the semi-metal category. Near the band-crossing point, known as the Dirac point, the energy-momentum dispersion is linear. This characteristic causes charge carriers to behave as massless relativistic Dirac fermions with a Fermi velocity only $300$ times smaller than the speed of light, drawing significant attention from physicists eager to explore relativistic physics in a solid-state material \cite{neto2009electronic}.

The symmetry of the valence and conduction bands around the Dirac point also results in the ambipolar nature of graphene, allowing the charge carrier type to be switched by applying a slight perturbation to the Fermi level around the Dirac point, for example, via an applied electric field, without the need for chemical doping. This symmetry also leads to electrons and holes sharing the same Fermi velocity, which has significant implications for the design of graphene-based pn-junction devices in electronics applications. The absence of backscattering in the graphene lattice leads to exceptionally high carrier mobilities, reaching up to $10,000$ cm$^2/$V-s in clean samples on Si/SiO$_2$ substrates at room temperature, and $250,000$ cm$^2/$V-s when sandwiched between hexagonal boron nitride (h-BN) layers \cite{novoselov_two-dimensional_2005,westervelt_graphene_2008,ponomarenko_chaotic_2008,orlita_approaching_2008}. 

As the thinnest known material with single-atomic-layer thickness, graphene has an exceptional surface-to-volume ratio, with the contact area of $sp^2$-hybridized carbons doubled. This high surface-to-volume ratio, combined with high carrier mobility at room temperature, makes graphene highly sensitive to its surrounding environment, rendering it an excellent material for sensor applications \cite{yavari_graphene-based_2012}. More importantly, graphene's ability to withstand considerable mechanical stress while maintaining its single-atomic-layer thickness renders it an ideal material for wearable devices and implants, such as skin-conformable electronic tattoos \cite{kireev2022continuous} and neural interfaces \cite{fabbro2016graphene}, which demand ultimate flexibility to conform to soft biological tissues seamlessly \cite{kuo_deformation_2013}. Graphene's potential in wearable and implantable biomedical applications is further reinforced by its biocompatibility \cite{chen_mechanically_2008, park_biocompatible_2010, yang_vivo_2011}.

Despite being only one atom thick, SLG displays notable opacity, absorbing $2.3\%$ of light across a wide spectrum, ranging from visible to infrared, due to its unique electronic structure \cite{nair_fine_2008}. Graphene's properties also lend themselves to plasmonic applications, attributable to its high carrier mobility \cite{novoselov_two-dimensional_2005}, tunability \cite{wang_gate-variable_2008}, and strong light confinement capabilities \cite{koppens_graphene_2011}. Chemically doped graphene offers potential in THz optoelectronics, as its transparency reduction can reach up to $40\%$ \cite{yan_infrared_2011}.

\subsection{Synthesis of GRMs}

\subsubsection{Top-down methods} Various top-down methods have been developed to produce graphene on different scales. Mechanical cleavage, one of the simplest methods, is employed to produce single- and few-layer graphene flakes from graphite via the repeated cleavage of graphite using a scotch tape \cite{novoselov_electric_2004}, as illustrated in Fig. \ref{fig:GRMSynthesis}(a). Chemical reduction of exfoliated GO is a solution-processed route for bulk-scale graphene production, commonly used for fabrication of organic devices and photovoltaic cells \cite{cote2009langmuir}. Another top-down method, liquid-phase exfoliation (LPE), entails exfoliating graphite in fluidic media using ultrasound waves to obtain single- or few-layer graphene (Fig. \ref{fig:GRMSynthesis}(b)). This process includes dispersion of graphite in a solvent, followed by exfoliation and purification steps \cite{monajjemi_liquid-phase_2017}. A relatively marginal yet interesting approach to producing graphene is unzipping CNTs to create graphene GNRs \cite{li_intercalation-assisted_2016} through various techniques, such as Argon plasma etching \cite{jiao_narrow_2009}, unwrapping by electrical current \cite{kim_graphene_2010}, and longitudinal unzipping of multi-walled CNTs \cite{higginbotham_lower-defect_2010}. 

\subsubsection{Bottom-up methods} 
\begin{figure}[t]
	\centering
	\includegraphics[width=1\columnwidth]{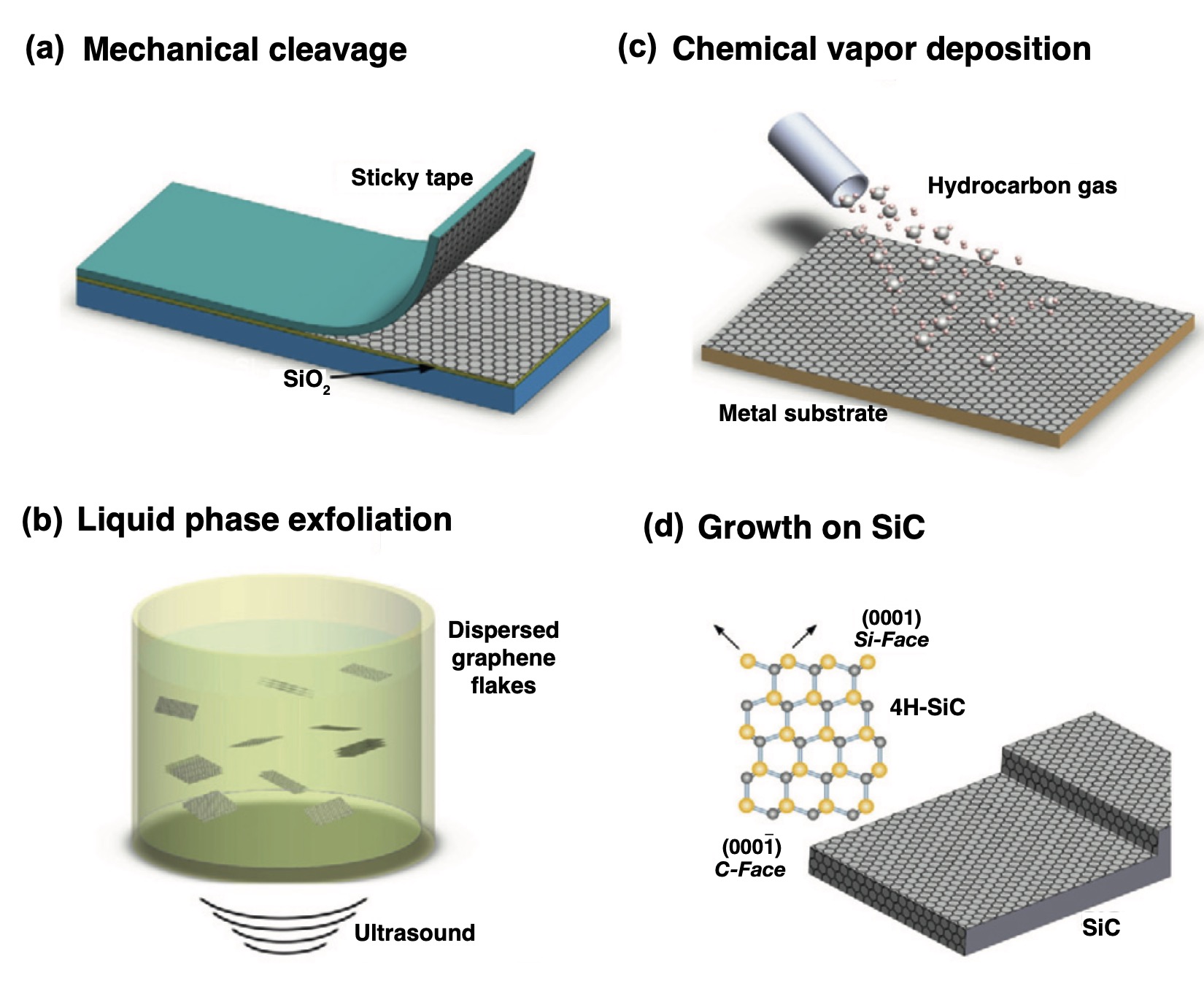}
	\caption{GRM synthesis methods: (a) mechanical cleavage, (b) liquid-phase exfoliation, (c) chemical vapor deposition, (d) epitaxial growth on SiC substrate. Reproduced with permission from \cite{bonaccorso2012production}.}
	\label{fig:GRMSynthesis}
\end{figure}

Chemical vapor deposition (CVD) is the most prevalent method for producing graphene, as it yields high-quality, large-area monolayers. In CVD, graphene is synthesized on metallic substrates, such as copper, in vacuum chambers supplied with gaseous precursors that react with the substrate at reaction rates set by chamber pressure and temperature (Fig. \ref{fig:GRMSynthesis}(c)). Graphene produced through this method is typically polycrystalline, and the grain boundaries in polycrystalline graphene can impair its electrical, mechanical, and biochemical properties \cite{yan_chemical_2014}. Moreover, the post-synthesis transfer of graphene from metallic substrates onto insulating surfaces, such as Si/SiO$_2$, required for electronic and sensing applications, introduces defects and impurities that can degrade device performance. Graphene can also be fabricated through epitaxial growth on silicon carbide (SiC) substrates (Fig. \ref{fig:GRMSynthesis}(d)). This method can produce high-quality, large-area graphene with uniform thickness, without the need for transferring it to another substrate, which is desirable for defect-free fabrication of electronic devices \cite{yazdi_epitaxial_2016}.

In summary, the choice of synthesis method can significantly impact the properties of the resulting GRMs, which, in turn, influences the performance of GRM-based devices. While mechanical cleavage can deliver the highest-quality graphene in terms of electrical and optical properties, this method is not scalable and exhibits low level of reproducibility. Conversely, CVD offers scalability for mass-producing large-area graphene; however, achieving single-crystal graphene, which is highly desired for sensing applications, is challenging with this method. Lastly, since LPE provides limited control over the quality and layer structure of the resulting graphene, its products are more suitable for use coating agents in printed electronics applications that typically do not demand high charge carrier mobilities.  

\section{GRM-based Micro/Nanoscale Transceivers}
\label{Sec:3}
The application environments of IoBNT, that primarily involve biological media, have been relatively unexplored by the ICT researchers until recently, as EM waves, which underpin conventional communication systems, are ill-suited for these settings. Hence, the exploration of unconventional communication approaches is of paramount importance to address the challenges posed by the unique characteristics of such environments. Among the emerging unconventional communication methods, MC, THz-band EM nanocommunications, and ultrasonic nanocommunications have garnered significant attention. 

These communication modalities exhibit significant deviations from conventional methods with respect to the form of information carriers, noise characteristics, achievable transmission rates, and communication delays. Moreover, in the context of THz-band EM and ultrasonic nanocommunications, although these modalities rely on wave propagation similar to conventional systems, EM and acoustic waves experience distinct physical phenomena at these operating frequencies in addition to complexities introduced by the biological environments. Consequently, new design approaches are essential for the practical implementation of Tx/Rx architectures tailored to the unique features of these unconventional modalities and the intricate biological settings. 

GRMs hold immense promise as a key enabler for realizing these novel communication methods. In this section, we present the state-of-the-art GRM technologies that have the potential to inspire the design of micro/nanoscale Tx/Rx components, encompassing antennas, signal sources, and detectors for MC, THz-band EM and ultrasonic nanocommunications in IoBNT applications. We highlight the key challenges in implementing full-fledged micro/nano Tx/Rxs and discuss the prospects of GRM-enabled novel Tx/Rxs in supporting multi-modal communications within the broader IoE framework.  

\subsection{Transceivers for Molecular Communications}
\label{MC}
Bio-inspired MC has emerged as the most promising communication method to realize the IoBNT. Capitalizing on molecules for information transfer, MC is inherently biocompatible, energy-efficient, and robust in physiologically-relevant environments such as within the human body. While there have been significant theoretical advances in understanding the performance limits of MC, such as channel capacity \cite{kuscu2019transmitter}, practical implementations of MC-capable artificial devices remain in their infancy.The radical differences between MC and conventional communication methods necessitate novel approaches for the physical design of MC Tx/Rxs. Recent demonstrations of prototype micro/nanoscale MC systems based on GRMs have revealed the enormous potential of nanomaterials in the development of practical MC devices \cite{kuscu2021fabrication}. 

As IoBNT applications, such as continuous health monitoring, are expected to closely interact with biological systems and tissues, it is imperative to ensure the biocompatibility and bio-durability of the materials used. This has led to the exploration of nanomaterials-based approaches for the design of MC-Tx/Rxs, which focus on chemically stable and biocompatible materials for sensing and releasing molecules, such as GRMs, CNTs, and silicon nanowires (SiNWs) \cite{kuscu2019transmitter}. While synthetic biology provides an alternative for designing biocompatible MC-Tx/Rxs \cite{unluturk2015genetically, egan2023toward}, it faces challenges concerning applicability and integration into functional systems \cite{soldner2020survey, liu2017using}. Thus, in the following, we will review the potential of nanomaterials, particularly GRMs, in designing and implementing practical MC-Tx/Rxs within the IoBNT framework.

\begin{figure}[t]
	\centering
	\includegraphics[width=1\columnwidth]{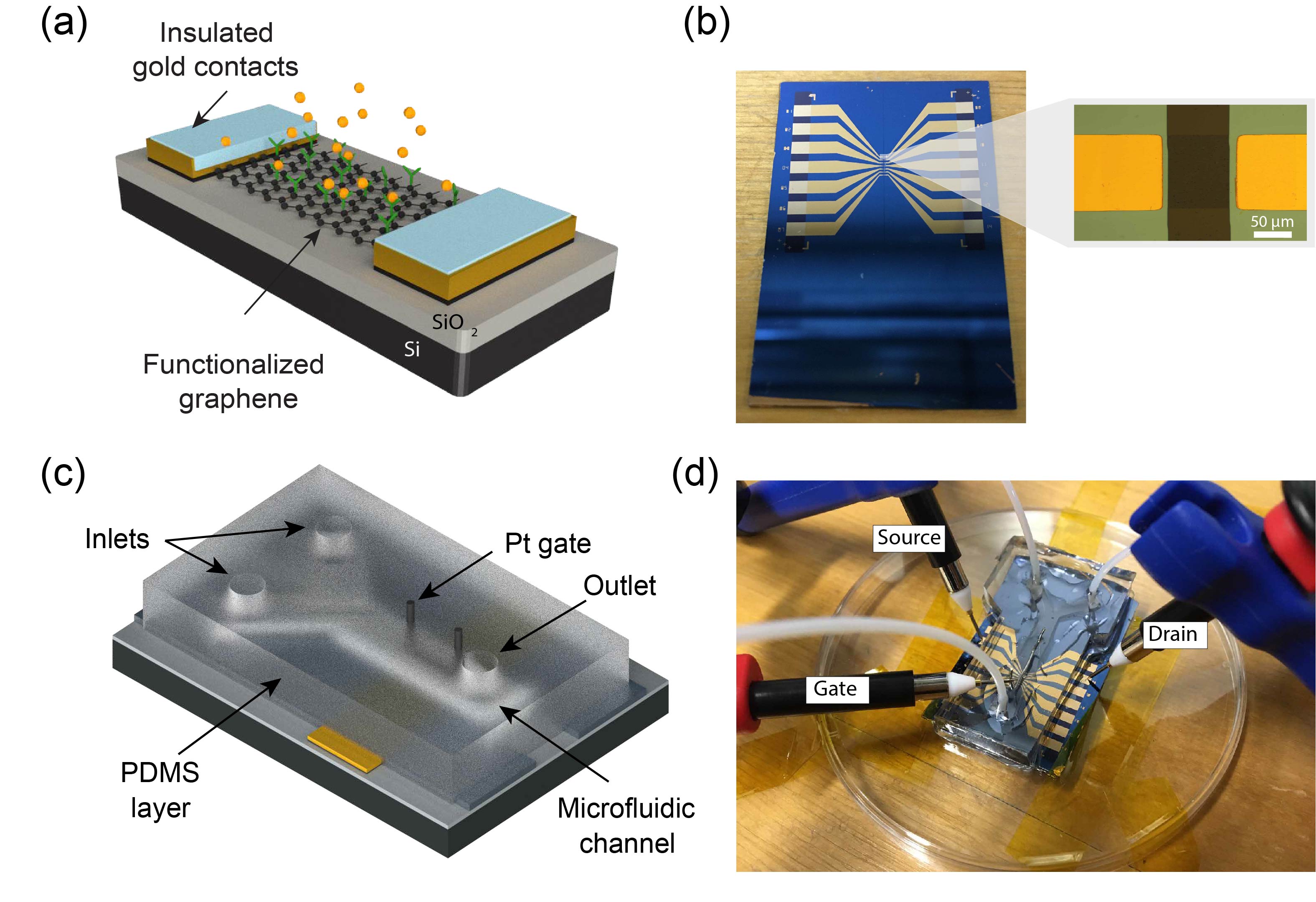}
	\caption{Graphene bioFET-based MC receiver {\cite{kuscu2021fabrication}: (a) Conceptual drawing of MC receiver, (b) Fabricated MC receiver, (c,d) Integration into a microfluidic chip.  } }
	\label{fig:MCreceiver}
\end{figure}

\subsubsection{\textbf{Molecular Communication Receiver}}

Drawing on the shared design principles between MC receivers (MC-Rxs) and biosensors, it is prudent to explore existing biosensor technologies, particularly those incorporating GRMs, as a foundation for designing MC-Rxs. Graphene has attracted significant attention in biosensing applications due to its remarkable properties, such as high charge mobility, chemical stability, and high sensitivity to biomolecules. Furthermore, graphene's 2D planar structure enables functionalization of its surface with an extensive array of biomolecular probes or receptors, facilitating selective sensing of various types of target molecules.

Among the diverse sensing modalities, including electrical, optical and mechanical, the FET-based biosensors (bioFETs) stand out for their ability to deliver miniaturization and high sensitivity, owing to their inherent signal amplification capability through the electrical field-effect \cite{economou2018flexible}. Graphene-based bioFETs, employing graphene as the transducer channel, have been reported to exhibit high sensitivity with detection limits in the sub-femtomolar range \cite{beraud2021graphene}, as well as rapid, real-time detection of various biomolecules \cite{seo2020rapid}. Although one-dimensional (1D) nanomaterials, such as SiNWs \cite{chen2011silicon} and single-walled CNTs \cite{curreli2008real}, have been widely considered for the transducer channel of bioFETs due to their comparable sensing performance to graphene-based transducers, challenges in their fabrication and scalable integration into existing complementary metal-oxide-semiconductor (CMOS) fabrication processes impede the development of 1D bioFETs \cite{tran2018cmos, beraud2021graphene}.

Graphene bioFETs also enable label-free, continuous molecular detection through the non-covalent functionalization of the graphene transducer channel with biological probes that reversible bind to the corresponding target molecules, such as aptamers and proteins, 
\cite{kuscu2019transmitter}. This capability has been demonstrated in a variety of biosensing applications, including the detection of pathogens, such as viruses and bacteria  \cite{seo2020rapid, huang2011graphene}, disease biomarkers for cancer and Alzheimer’s Disease \cite{tsang2019chemically, bungon2021graphene}, and nucleic acids \cite{hwang2020ultrasensitive}. Inspired by the potential of graphene bioFETs, Kuscu \emph{et~al.} recently introduced the first micro/nanoscale MC-Rx based on graphene bioFETs, employing single-stranded DNA molecules as information carriers and their complementary DNA molecules as receptors on graphene transducer surface, as illustrated in Fig.~\ref{fig:MCreceiver} \cite{kuscu2021fabrication}. 

\begin{figure*}[!t]
	\centering
	\includegraphics[width=1\textwidth]{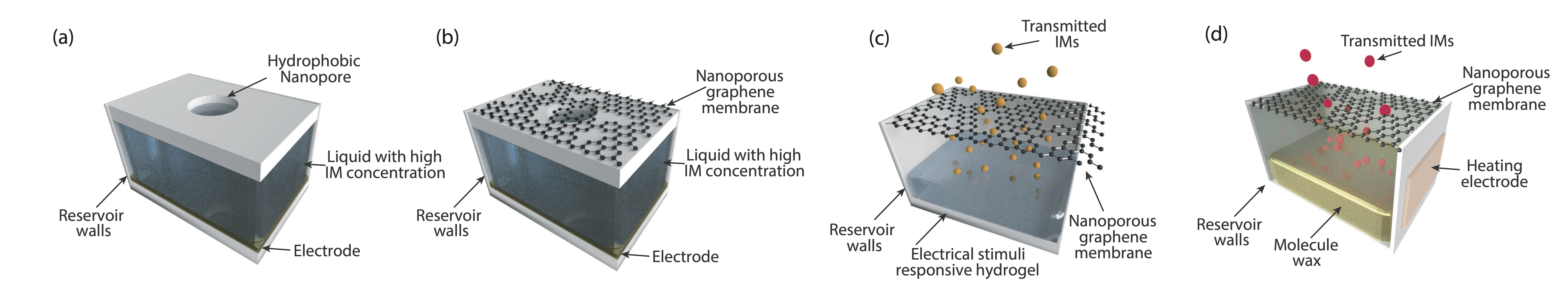}
	\caption{Conceptual MC-Tx architectures \cite{kuscu2019transmitter}: (a) Liquid-borne MC-Tx with hydrophobic nanopore. (b) Liquid-borne MC-Tx with hydrophobic nanopore and nanoporous graphene membrane. (c) MC-Tx with hydrogel and nanoporous graphene membrane. (d) Airborne MC-Tx with molecule wax and nanoporous graphene membrane.} 
	\label{fig:MCtransmitter}
\end{figure*}

Practical applications of graphene bioFET-based MC-Rxs currently face several challenges, including issues relatd to reproducibility and reliability. CVD growth of graphene on copper substrates is the most common method, but the introduction of defects and impurities during the graphene transfer process onto another substrate for bioFET fabrication is unavoidable. Si/SiO\textsubscript{2} substrates are typically used in bioFETs due to their compatibility with the conventional microfabrication processes. However, these substrates can introduce charged impurities that reduce the charge carrier mobility of CVD-grown graphene in biosensing applications \cite{chen2008intrinsic}. Consequently, alternative substrates or growth methods, such as direct graphene growth on non-metallic substrates, such as organic flexible materials \cite{xu2018ultrasensitive}, \cite{economou2018flexible}, warrant further investigation. Moreover, previse control over the coverage, stability, and orientation of immobilized probes on the graphene transducer channel is still lacking, \cite{kuscu2016physical,beraud2021graphene}, which must be addressed to resolve reproducibility concerns. 

Additionally, the potential interaction of graphene with non-specific molecules present in the communication channel poses a challenge for the reliable operation of bioFET-based MC-Rxs in biological environments, where numerous types of molecules and ions can co-exist. For example, in physiological environments with high ionic concentrations, the molecular detection performance of MC-Rxs may be significantly degraded due to the Debye screening effect. Researchers are exploring novel strategies to mitigate these phenomena, such as encapsulating the bioFET with a biomolecule-permeable polymer layer to increase the effective screening length \cite{gao2015general, beraud2021graphene} or employing surface passivation techniques to minimize non-specific interactions \cite{zhang2020ultrasensitive}.

\subsubsection{\textbf{Molecular Communication Transmitter}}
MC transmitters~(MC-Txs) encode information into various physical properties of molecules, such as concentration and type, before releasing modulated molecular signals into the MC channel. Practical implementation of MC-Txs remains an open issue, as there is a lack of literature on precise, controllable molecule delivery structures and micro/nanoscale molecule reservoirs capable of sustaining continuous molecule release. GRMs can offer significant advantages in addressing these challenges, due to their unique electronic, mechanical, and chemical properties.

Kuscu \emph{et~al.} proposed micro/nanoscale MC-Tx architectures for both liquid-borne and airborne MC communications, utilizing nanoporous graphene membranes to cap molecule reservoirs \cite{kuscu2019transmitter}, as depicted in Fig \ref{fig:MCtransmitter}. The permeability of the nanoporous graphene membrane is regulated by an electric field, facilitating selective and tunable release of information-carrying molecules (IMs) stored in the reservoir, while minimizing molecule leakage, a challenge associated with hydrophobic nanopores, as shown in Fig.~\ref{fig:MCtransmitter}(a). The use of graphene membranes offers several advantages over conventional membranes, such as high mechanical strength, large surface area, low thickness, and tunable pore size \cite{kuscu2019transmitter}. 

In the liquid-borne MC-Tx architecture illustrated in Fig.~\ref{fig:MCtransmitter}(b), an electrode plate at the bottom of the molecule container provides further control over molecule release by attracting or repelling charged molecules using the electric field. The architecture shown in Fig.~\ref{fig:MCtransmitter}(c) employs electrically-responsive hydrogels as molecule reservoirs, enabling modulation of molecule release through electrical stimulation. The airborne MC-Tx architecture in Fig.~\ref{fig:MCtransmitter}(d) incorporates a wax layer to store molecules, which are continuously released when heated by an electric field.
Key concerns for these architectures include molecular leakage (non-negligible molecule release without electrical stimuli), and transmission delay (time required for information molecules to enter the MC channel after electric field application). GRMs can help mitigate these issues by providing a barrier layer (graphene membrane) between the reservoir and the environment, and reinforce hydrogels or polymeric waxes with improved mechanical properties and responsiveness \cite{lazuar2023graphene}.


\subsection{Transceivers for THz-band Nanocommunications}
\label{THz}
As IoBNT applications involve biological environments such as tissues, unique challenges arise in various aspects, necessitating the transformation of material and device technologies to accommodate the unique requirements of these environments. Conventional EM communication technologies, namely radio frequency (RF) technologies, are not suitable for IoBNT due to the limited miniaturization of RF Tx/Rx components at the required operating frequencies and power levels, which exceed the capabilities of BNTs. However, the emerging field of THz-band communications, employing high-frequency EM waves in the underexplored $0.1$ to $10$ THz range, appears promising for IoBNT, as it facilitates micro/nanoscale miniaturization and low-power operation. Moreover, THz-band communications exhibits biocompatibilty, as THz radiation does not ionize biological tissues, lacking sufficient energy for ionization of atoms and molecules \cite{zhao2014advances}. 

Since the THz-band lies between microwave and infrared frequencies, THz waves exhibit both photonic and electromagnetic properties, rendering electronic and photonic device technologies applicable for THz Tx/Rx design. On the other hand, THz device technology remained underdeveloped until recently, primarily due to the scarcity of practical signal generators and detectors for the so-called THz gap. More specifically, conventional solid-state electronic devices generate limited signal power at THz frequencies, while photonic devices have been constrained by the absence of materials with suitable electrical and optical properties \cite{chaves2020bandgap}. Nonetheless, the emergence of GRMs with exceptional electronic and optical properties has significantly propelled the advancements in THz device technologies.

Graphene, with its gapless band structure, demonstrates a spectral response at THz frequencies, unlike other 2D materials with larger band gaps \cite{liu2018towards}. This attribute enables EM radiation-graphene interaction at THz frequencies, making it suitable for THz emission, modulation, and detection. Graphene's electrical, optical and plasmonic properties can be tuned through chemical doping or electrical/magnetic biasing \cite{correas2017graphene}, allowing the development of reconfigurable THz Tx/Rx components, such as nano-antennas with dynamic radiation patterns controlled by graphene's tunable surface conductivity \cite{andrello2018dynamic}. Moreover, the ease of integrating graphene-based THz components with each other, as exemplified by integrated modulators and detectors, could facilitate low-power, on-chip THz Tx/Rxs, highly sought after for standalone IoBNT applications. Additionally, graphene's low electronic noise temperature makes it suitable for physiological environments, preventing performance degradation in THz Tx/Rx devices \cite{yang2016channel}. 

\begin{figure*}[t]
	\centering
	\includegraphics[width=\textwidth]{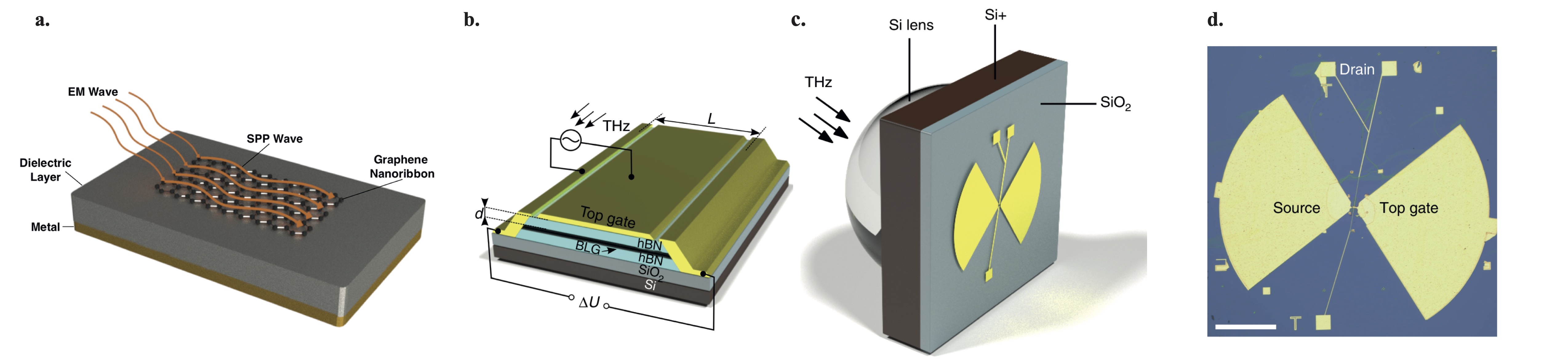}
    	\caption{(a) Conceptual design of a graphene nanoribbon-based plasmonic antenna for THz communication. (b, c, d) Graphene-based THz detector: (b) Bilayer graphene FET architecture of the detector, (c) 3D view, and (d) corresponding optical micrograph \cite{bandurin2018resonant}.}
	\label{fig:THz_comp}
\end{figure*}

Graphene's support for tunable surface plasmon polaritons (SPPs) at THz frequencies offers great potential for realizing micro/nanoscale THz Tx/Rxs. Confinement of THz waves in a sub-wavelength scale in the SPP mode enables high-order miniaturization \cite{andrello2018dynamic} due to the smaller wavelength of SPP waves compared to EM waves, resulting from the lower propagation speed of SPP waves on graphene \cite{jornet2010graphene}. Despite offering high miniaturization, plasmonic antennas based on pure graphene exhibit low radiation efficiency \cite{akylidiz2013graphene}. Radiation efficiency can be improved by employing substrate materials with a higher dielectric constant or increasing the number of graphene layers, as demonstrated by hybrid graphene-dielectric structures \cite{hosseininejad2016surveying} and few-layer graphene stacks \cite{hosseininejad2018reconfigurable}. However, this compromises the level of miniaturization \cite{ hosseininejad2016surveying}. 
Consequently, further research is necessary to address the trade-off between miniaturization and radiation efficiency.

The utilization of graphene SPPs enables frequency-reconfigurable THz plasmonic antennas \cite{ akylidiz2013graphene} and modulators \cite{singh2016graphene, nafari2018plasmonic}, paving the way for efficient, integrated on-chip all-graphene THz Tx/Rxs \cite{jornet2014graphene} along with on-chip plasmonic THz signal sources \cite{crabb2021chip}, and detectors \cite{bandurin2018resonant} as exemplified by the THz plasmonic components shown in Fig.~\ref{fig:THz_comp}. The frequency-reconfigurability of graphene sheets is achieved by varying the chemical potential (Fermi level) of graphene, typically via electric field application. Altering the chemical potential allows control of graphene surface conductivity, thus affecting the propagation constant and resonance frequency of graphene SPPs. For example, increasing the chemical potential shifts the resonant frequency upwards, resulting in higher radiation efficiency \cite{lallas2019key}. However, it should be noted that electronic reconfigurability is adversely affected by high chemical potentials \cite{correas2017graphene}. 

Several GRM-based configurations have been reported in the literature that can simultaneously provide high radiation efficiency and frequency reconfigurability, such as tunable THz metamaterials comprising metallic resonance structures with embedded graphene \cite{luo2019graphene}, and multiple graphene sheets controlled by different bias voltages \cite{zhang2021design}. Nevertheless, the feasibility of these configurations for IoBNT applications must be evaluated in terms of miniaturization. Additionally, complex physiological environments with various co-existing molecules can present significant challenges for the reliable operation of plasmonic Tx/Rx components. Specifically, the adsorption of interfering molecules on the graphene surface can alter its refractive index, subsequently affecting the propagation characteristics of the SPPs \cite{rodrigo2015mid}. Therefore, a comprehensive analysis of interfering molecular interactions with graphene under different physiological conditions is required to assess their impact on graphene's reconfigurability.

\subsection{Transceivers for Ultrasonic Nanocommunications}
\label{Acoustic}
Ultrasonic nanocommunications utilizing high-frequency acoustic waves (above $20$ kHz) represents a promising approach for IoBNT, particularly in intra-body applications \cite{hogg2012acoustic}. Acoustic communications has certain advantages over EM-based communications in water-dense media. For example, sound is often preferred for underwater communication as RF waves experience significant absorption losses. Therefore, acoustic communication within the body, where water is the primary constituent, may also be advantageous.

Moreover, ultrasonic nanocommunications has the potential to support high data rates in intra-body applications. In one study, researchers experimentally demonstrated ultrasonic propagation through a tissue-mimicking environment, achieving data rates up to $700$ kbps with a $40 \mu m$ transmit power \cite{santagati2016experimental}. However, ultrasonic Tx/Rxs must be tailored for IoBNT in several aspects, including size, robustness, and biocompatibility. This is because BNTs have micro/nanoscale dimensions, propagation in biological environments with numerous obstacles requires robustness, and transmission in biological tissues must adhere to safety limits concerning acoustic pressure and power.
To meet these requirements, directional transmission at high frequencies (around hundreds of MHz) is necessary for \textit{in vivo} communications \cite{hogg2012acoustic}. 
Additionally, wideband ultrasonic nanocommunications employing short pulses with low duty cycles can provide compatibility with IoBNT by reducing thermal and mechanical effects on tissues and offering robustness against the multipath fading caused by obstacles inside the body \cite{santagati2016experimental}.


The primary component in ultrasonic communications is the ultrasonic transducer, which emits and detects ultrasound waves. Conventional ultrasonic transducers typically rely on the piezoelectric effect, converting mechanical vibrations to electrical signals and vice versa. However, these transducers often operate only near their resonance frequency and 
allow only limited miniaturization due to performance and fabrication constraints \cite{wissmeyer2018looking}. Alternatively, optoacoustic generation of acoustic waves is a powerful method for generating high-frequency, wideband ultrasound. Using light-absorbing nanomaterials providing efficient light-to-ultrasound conversion, such as GRMs and CNTs, miniaturized ultrasonic transmitters can be developed. These transmitters can be integrated with wideband sensitive, all-optical detectors, such as optical interferometric detectors, enabling the implementation of all-optical, wideband, high-frequency ultrasound transducers with strong miniaturization potential \cite{ wissmeyer2018looking}.

\begin{figure*}[!t]
    \centering
    \includegraphics[width=\textwidth]{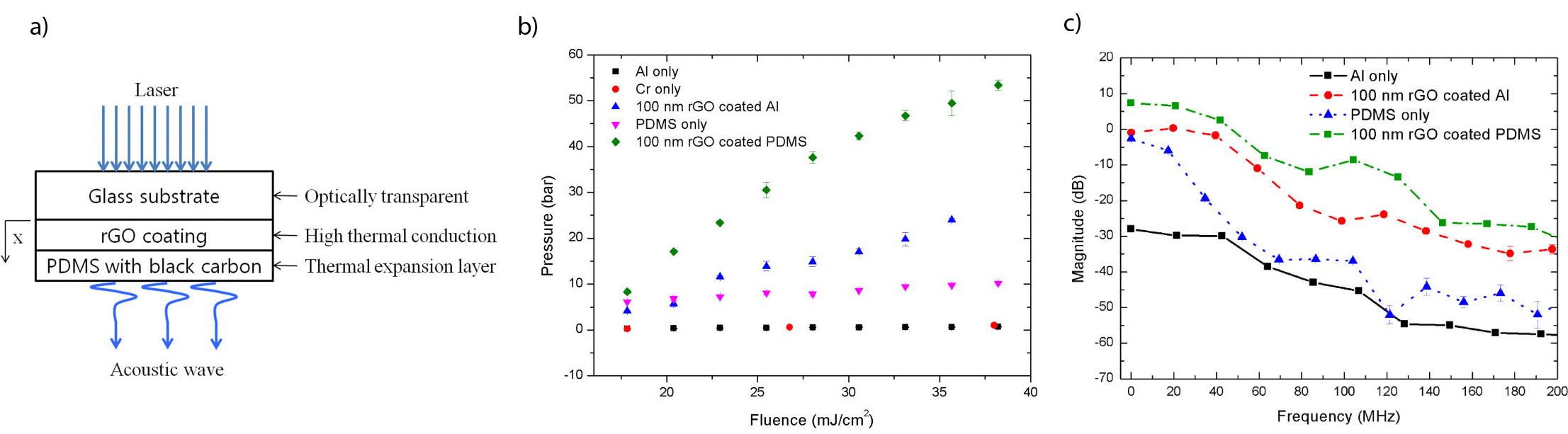}
    \caption{Optoacoustic ultrasound transmitter based on rGO-coated PDMS film: (a) Schematic of the film. (b) Maximum pressure of ultrasonic waves from different samples at various fluences. (c) Frequency spectra of ultrasonic waves for different samples. Reproduced from \cite{lee2015reduced}, with the permission
of AIP Publishing.}
    \label{fig:ultrasonic}
\end{figure*}

The photoacoustic effect describes the generation of sound waves via modulated optical waves. The absorbed light causes the material to heat up, leading to sound emission via thermoelastic expansion. As a result, wideband ultrasound can be generated using short laser pulses on light-absorbing materials. For \textit{in vivo} ultrasonic transmitters, efficient light-to-sound conversion is crucial to achieve adequate acoustic pressure, considering the limited laser pulse energy to avoid harmful biological effects. Conversion efficiency mainly depends on the absorbing material, which should ideally possess high optical absorption capacity, high thermal conductivity, and low heat capacity \cite{giorgianni2018high}. Carbon-based nanomaterials, such as GRMs and CNTs, with their exceptional photothermal properties and wideband light absorption capability, are well-suited for efficient photoacoustic conversion. Both rGO and CNT have been extensively explored for optoacoustic ultrasound generation \cite{ lee2015reduced}. In biological environments, rGO is favored over CNT due to its larger surface area and improved dispersibility \cite{fu2019photoacoustic}. 

Enhancing photoacoustic conversion efficiency is commonly achieved through engineering composite nanomaterials. For instance, rGO can be combined with metallic materials, such as gold nanoparticles (AuNPs), to improve its light-absorbing capability \cite{moon2015amplified}. Additionally, rGO can be mixed with metals and polymers acting as thermal expansion layer, such as aluminum (Al) and polydimethylsiloxane (PDMS), for enhanced thermoelastic expansion and increased acoustic pressure \cite{hwan2012reduced, lee2015reduced}.  Comparatively, rGO-PDMS ultrasound transmitters exhibit higher photoacoustic pressure than rGO-Al, as PDMS has greater thermoelastic expansion than Al \cite{lee2015reduced}, as demonstrated in Fig.~\ref{fig:ultrasonic}. Optimizing composite material properties for maximum sound pressure requires considering certain trade-offs. For examples, while rGO's light absorptivity increases with its thickness \cite{hwan2012reduced}, the bandwidth of the generated ultrasonic pulses decreases with the thickness of the composite \cite{lee2018efficient}.

Piezoelectric detectors are commonly used for detecting ultrasonic signals in optoacoustic imaging. However, these detectors have an inherently narrow operation frequency, and their sensitivity decreases as the detector size is reduced \cite{wissmeyer2018looking}. In contrast, optical interferometric resonators are not size-dependent and can be miniaturized. All-optical generation and detection of ultrasound can enable on-chip integrated optical ultrasonic Tx/Rx, a direction further motivated by advancements in nanolaser technology, such as plasmonic nanolasers \cite{azzam2020ten}. Nevertheless, fabricating on-chip integrated plasmonic nanolasers presents significant challenges, including the need for electrical pumping without harming the biological environment. To address this issue, it is crucial to lower the pumping threshold and maintain high efficiency \cite{liang2020plasmonic}.

\color{black}

\subsection{Multi-modal Transceivers for IoBNT}
As discussed in previous sections, GRMs can enable the reliable and information-efficient transduction of diverse communication modalities into the electrical domain and vice versa. This inter-modality transduction property of graphene can be harnessed to develop Tx/Rxs capable of seamlessly interfacing heterogeneous networks within IoBNT, and connecting IoBNT with the conventional communication networks.

We recently introduced a device concept called \textit{universal transceivers} \cite{civas2021universal}, characterized by \textit{multi-modality} (in communication and energy harvesting), \textit{modularity}, \textit{tunability}, and \textit{scalability} (to micro/nanoscales). Universal Tx/Rx architectures possessing these attributes can address the major inter-operability challenges in realizing the emerging \textit{Internet of Everything (IoE)} paradigm. This paradigm has the potential to revolutionize our interaction with the physical world \cite{civas2021universal} by integrating various Internet of Things (IoT) technologies and applications, including IoBNT. 

However, device complexity poses a significant challenge for the physical implementation of universal Tx/Rxs, as each communication modality adds to the overall hardware complexity. Direct inter-modality transduction without conversion to the electrical domain, where signals are processed, can considerably reduce device complexity. For example, molecule-sensitive graphene field-effect transistors (GFETs) capable of simultaneously sensing molecules and modulating the output RF signal have been reported \cite{hajizadegan2017graphene}. GRMs with multi-functional capabilities in other forms also show promise in addressing device complexity, as demonstrated by a multi-functional graphene-based modulator combining storage and transduction capabilities \cite{huang2016chemical}.

\section{GRM-based Bio-cyber Interfaces}

\begin{figure*}[t]
	\centering
	\includegraphics[width=0.8\textwidth]{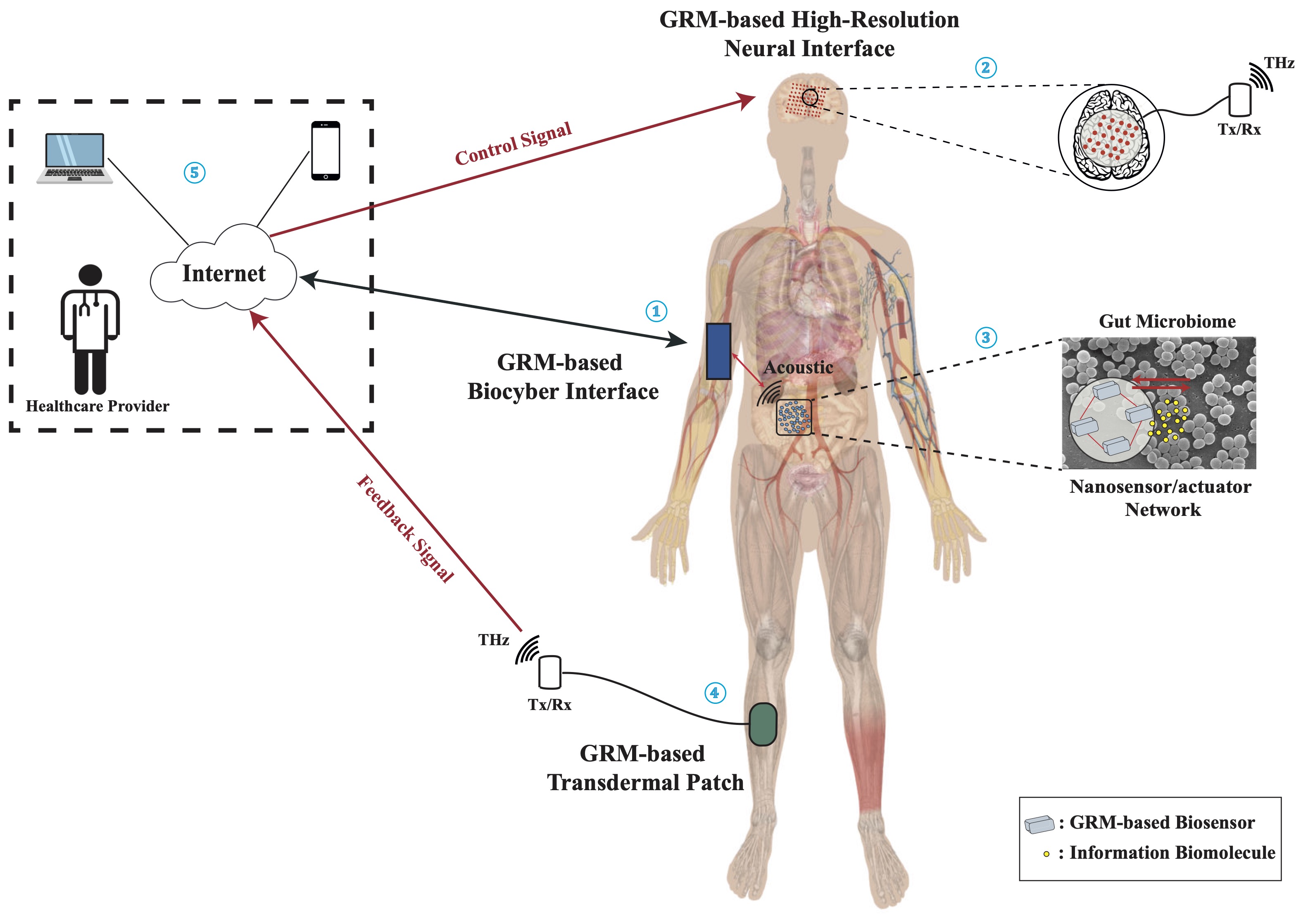}
    	\caption{An IoBNT architecture with GRM-based bio-cyber interfaces facilitating bi-directional communication between intrabody nanonetworks and conventional networks: \\
(1) GRM-based bio-cyber interface enabling bi-directional communication between the Internet and intra-body networks. \\
(2) GRM-based high-resolution neural interface connected to external networks via THz signals. \\
(3) Nanosensor/actuator network monitoring and controlling gut microbiome composition, connected to a bio-cyber interface through acoustic signals. \\
(4) GRM-based transdermal patch providing sensory feedback to (2) via the Internet by converting biochemical signals to THz signals and transmitting them to the Internet infrastructure for further processing. \\
(5) The cyber domain of the Internet. }
	\label{fig:biocyber}
\end{figure*}

Realizing the envisioned IoBNT applications often necessitates bridging the gap between bio-nano networks, which typically operate in the biochemical domain (e.g., intra-body), and conventional networks that primarily rely on electromagnetic signal exchange. The disparities between these heterogeneous networks in terms of communication media, modality, and compatible device technologies require sophisticated interfaces capable of seamlessly interconnecting these networks by overcoming interoperability challenges (Fig. \ref{fig:biocyber}). These so-called bio-cyber interfaces have been the focus of extensive research, with various innovative methods proposed in the IoBNT context, including those based on optogenetics \cite{balasubramaniam2018wireless}, THz control of protein conformations \cite{elayan2020information}, stimuli-responsive hydrogels \cite{kuscu2019transmitter}, redox electrochemisty \cite{vanarsdale2020redox}, and biosensing \cite{kuscu2016physical, kuscu2016modeling}. These methods offer uni-directional or bi-directional bio-cyber interfacing with varying spatiotemporal resolution. In this section, we provide a brief overview of IoBNT bio-cyber interfacing technologies where GRMs can provide unique advantages.

\subsection{GRM-based Biosensors for IoBNT}

Depending on the specific IoBNT application, biosensors can serve as bio-cyber interfaces, converting molecular signals into electrical or other signal forms that can be processed and transmitted over conventional communication networks. The use of biosensors as bio-cyber interfaces within IoBNT applications, however, presents challenges related to size, energy consumption, biocompatibility, and the establishment of reliable bi-directional communication.

A wide array of IoBNT applications require interfacing with living systems and tissues, which necessitates biosensors at the micro/nanoscale and demands high-level biocompatibility. This is crucial for preventing toxic and adverse effects on living cells and biochemical processes therein, while also ensuring the proper operation of biosensors without performance degradation \cite{pramanik2020advancing,syama2016safety}.

\begin{figure}[!t]
    \centering
    \includegraphics[width=\linewidth]{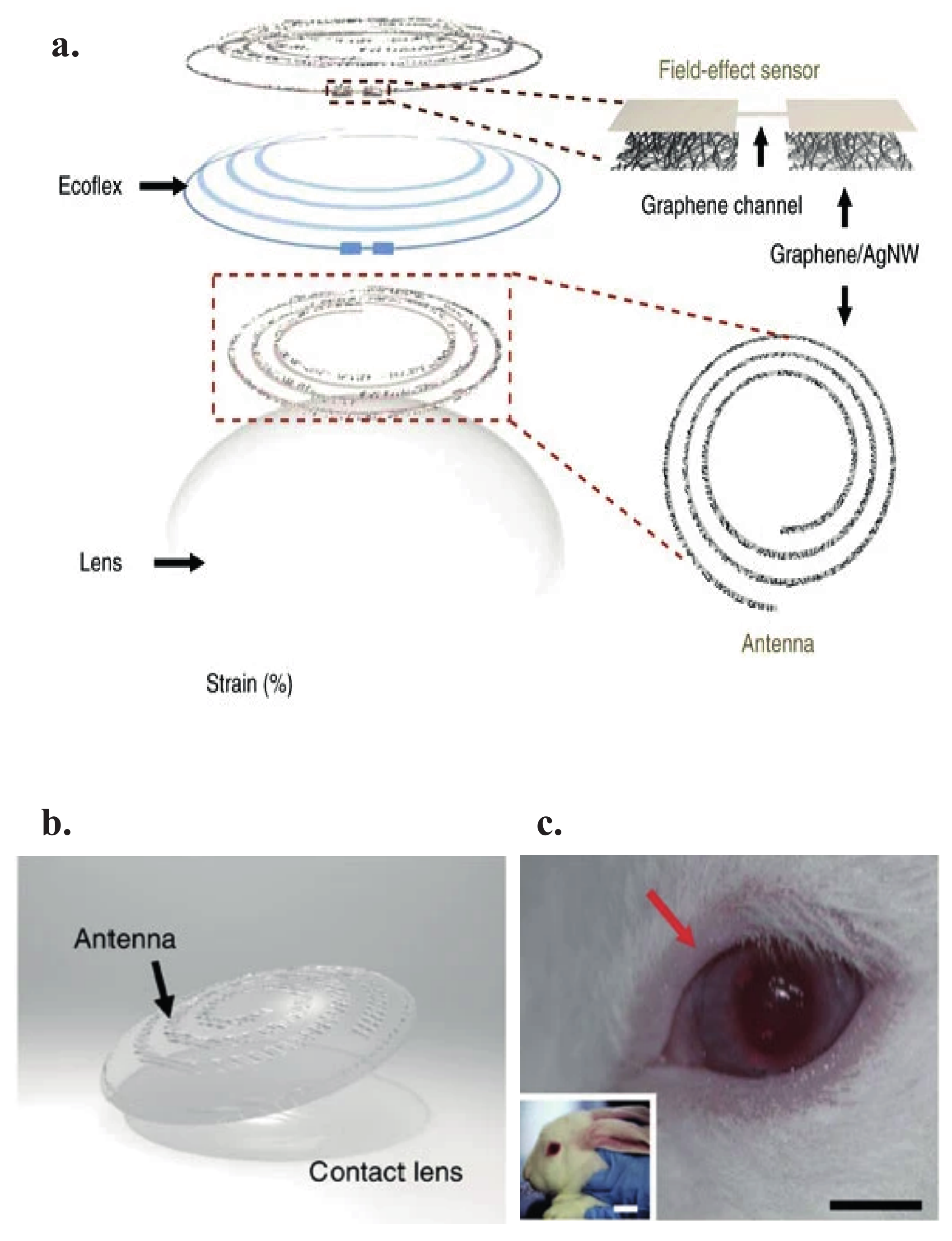}
    \caption{GRM-based wearable smart contact lens \cite{kim2017wearable}: (a) GRM-based wearable contact lens sensor integrating a
glucose sensor and an intraocular pressure sensor. (b) Illustration of the transparent glucose sensor on the contact lens. (c) GRM-based wireless sensor integrated onto a rabbit's eye. }
    \label{fig:smart_lens}
\end{figure}

The performance of a biosensor is directly related to its transduction process. Reliable and information-efficient inter-modality transduction (e.g., biochemical signals to electrical signals) is required to eliminate processing delays and information loss. Ensuring information-efficient and rapid transduction is particularly critical for healthcare applications of IoBNT, such as early diagnosis and treatment of diseases.

Biosensing is one of the areas where GRMs excel due to their biocompatibility, flexibility, high charge carrier mobility, and unique 2D structure with a significantly high surface-to-volume ratio, leading to exceptional sensitivity \cite{celik2015graphene,liao2018graphene,holzinger2014nanomaterials,szunerits2018graphene,pinto2013graphene}. Their capacity to be functionalized with a diverse array of bio-recognition elements, such as proteins, enzymes, and DNA/RNA aptamers, makes them versatile platforms for selective interfacing with various biochemical systems and signals. Notably, aptamer-functionalized graphene sensors have demonstrated real-time monitoring of cytokines in human biofluid, achieving highly sensitive detection of disease biomarkers \cite{wang2020wearable}. Furthermore, GRM-based biosensors can transduce biochemical signals into a broad range of signal forms, such as electrical, optical, RF/microwave/THz electromagnetic signals \cite{ferrari2015science,civas2021universal}, making them ideal candidates for addressing bio-cyber interfacing challenges in IoBNT.

FET-based biosensors, also known as bioFETs, represent a widely adopted architecture for GRM-based biosensors due to their inherent signal amplification capability. In bioFETs, the GRM-based transduction channel of the transistor is in direct contact with the biochemical environment, with biorecognition elements, i.e., receptors, serving as the gate electrode. An example application of GRM-based bioFETs is wearable contact lens sensors (Fig. \ref{fig:smart_lens}), which detect glucose concentration in tears and wirelessly transmit data to a mobile device for blood glucose level estimation. \cite{kim2017wearable}. Exploiting correlations between biomarker levels in blood and those in bodily fluids, e.g., tears, saliva, and sweat, that are easily accessible by wearable GRM-based biosensors is crucial for bio-cyber interfacing in IoBNT applications, where the molecular signals generated by intra-body MC networks can be detected non-invasively. 

In conclusion, the advancements in GRM-based biosensors hold great potential for addressing the challenges associated with bio-cyber interfacing in IoBNT applications. These cutting-edge biosensors offer unprecedented sensitivity, versatility, and biocompatibility, opening new avenues for IoBNT applications in healthcare and beyond.


\subsection{GRM-based Neural Interfaces for IoBNT}

The nervous system is the most advanced and complex human-body network \cite{malak2014communication} with the intricacies of many of its signaling pathways yet to be unfolded. It is a very large-scale network distributed throughout the body for transferring external stimuli to the brain and \textit{vice versa}. The nervous system utilizes electrical and chemical signals to transfer information among its constituents, standing as a multi-modal and natural communication infrastructure for artificial IoBNT networks. In that regard, bi-directional interfacing with the nervous system is essential to enable many of the foreseen IoBNT applications.

One promising target for IoBNT applications is the bi-directional gut-brain axis, which connects the gut microbiota with the central nervous system. The electrochemical signaling pathways in this axis play a significant role in numerous neurological disorders, such as autism and depression, as well as digestive conditions, e.g., irritable bowel syndrome \cite{mayer2015gut,cryan2019microbiota,rhee2009principles,grenham2011brain}. Implementing an IoBNT system capable of bi-directional communication with the gut-brain axis through bio-cyber interfaces could facilitate monitoring and controlling gut-brain signaling for early diagnosis and treatment of associated diseases. Moreover, IoBNT networks interfacing with the nervous system could be employed to reestablish disrupted neural signaling between the brain and muscles in spinal cord injuries, ultimately restoring motor functions in paralyzed patients \cite{akan2021information}.

Interfacing with the nervous system is conventionally realized via electrical stimulation of the neurons and transduction of dynamic electrochemical neural activities into electrical signals \cite{zhang2020electronic,jackson2012neural,grill2009implanted}. Achieving high spatio-temporal resolution neural interfaces remains a significant challenge due to the complex anatomy of nervous tissues and intricate interconnections among neurons, leading to considerable noise and interference. These challenges are currently being addressed with advanced nanomaterials and flexible electronic technologies.

Exceptional physicochemical properties of GRMs, such as extraordinary charge carrier mobility, electrochemical stability, mechanical flexibility, miniaturization capacity, and transparency make them ideal for addressing the aforementioned challenges in neural interface design \cite{kostarelos2017graphene,huang2019graphene,royal2019ihuman}. GRM-based neural interfaces enable unprecedented access to the spatio-temporal dynamics of neural activity, facilitating remote monitoring and control of the nervous system. For instance, transparent graphene electrode arrays allow high-resolution visualization of cells during neural imaging and recording. In \cite{liu2018compact}, the authors propose using transparent graphene micro-electrodes to develop a compact closed-loop system for optogenetic stimulation, selectively exciting and inhibiting genetically targeted cells through photons. This optical control of genetically engineered cells enables external activation of light-sensitive ion channels, further modulating the release of specific information molecules, such as neurotransmitters, neuropeptides and ions. Beyond monitoring and controlling the nervous system's electrical activity, GRM-based neural interfaces also create opportunities for \textit{in situ} biosensing and smart drug delivery applications. Real-time detection of neurotransmitters and neuromodulators, e.g., serotonin, and monitoring ion channel activity can enable early diagnosis of neural disorders and the release of therapeutic molecules based on biosensing readouts \cite{kostarelos2017graphene}.

In conclusion, GRM-based interfacing holds immense potential as a bridge between communication engineering and life sciences, paving the way for the development of future IoBNT-based diagnosis and treatment applications for communication-related neural disorders.

\subsection{GRM-based Drug Delivery Systems for IoBNT}

Stimuli-response smart drug delivery systems hold significant potential for informing the design of bio-cyber interfaces in IoBNT applications. These advanced technologies can precisely modulate molecule release rates in response to time-varying external stimuli, such as electrical, optical, and magnetic signals. This capability enables seamless interfacing between conventional communication networks and bio-nano networks that interact through molecular communications. 

GRMs have emerged as valuable components in the development of stimuli-responsive drug delivery systems. Their unique properties, such as hydrophilicity, high surface area, and ability to form composites, make them ideal candidates for this purpose. Specifically, GO composites with various types of nanoparticles have been widely explored for their capacity to respond to near-infrared (NIR) light and magnetic-field stimuli for controlled molecule release \cite{yang2016stimuli}. These composites exhibit excellent biocompatibility, stability, and tunable physicochemical properties, leading to highly effective and targeted drug delivery systems. 

Another promising application of GRMs in drug delivery is the development of electro-responsive hydrogels \cite{reina2017promises}. These hydrogels, when combined with graphene, can exhibit enhanced drug release performance by providing reliable and precise control over molecule release through the magnitude, duration, and frequency of the applied field. Although activating drug delivery devices via an electrical field may cause resistive heating problems; the incorporation of pristine graphene with an electro-sensitive hydrogel matrix mitigates this issue. This combination results in a mechanically and chemically stable interface, suitable for triggering drug release through electrical fields \cite{servant2014graphene}. 

Further advances in GRM-based drug delivery systems could lead to more sophisticated and responsive bio-cyber interface designs for IoBNT applications. For instance, integrating multi-stimuli responsive components, such as thermo-responsive polymers, could enable even more precise control over drug release profiles, increasing the bandwidth of transmitted molecular signals. Additionally, combining GRMs with other nanomaterials, such as gold nanoparticles or quantum dots, can facilitate multi-modal drug delivery systems that simultaneously support imaging, sensing, and therapeutic functions, enabling bi-directional bio-cyber interfaces for IoBNT.

\section{GRM-based Micro/Nanoscale Energy Harvesting and Storage}

Ensuring uninterrupted sensing, computation, and data communication at the micro/nanoscale requires efficient utilization of limited energy resources and the adoption of self-sustaining strategies, such as energy harvesting (EH), to fulfill the stringent requirements of IoBNT applications. However, the projected scale and capabilities of the components required for this purpose present challenges to existing technologies, prompting the development of novel and unconventional solutions. GRMs have recently garnered significant interest in this field due to their exceptional properties. In this section, we examine the potential of GRMs in advancing micro/nanoscale energy harvesting and storage technologies, which will ensure the perpetual operation of IoBNT applications.


\subsection{GRM-based Micro/Nanoscale Energy Harvesting for IoBNT}

\begin{figure}[!t]
    \centering
    \includegraphics[width=\linewidth]{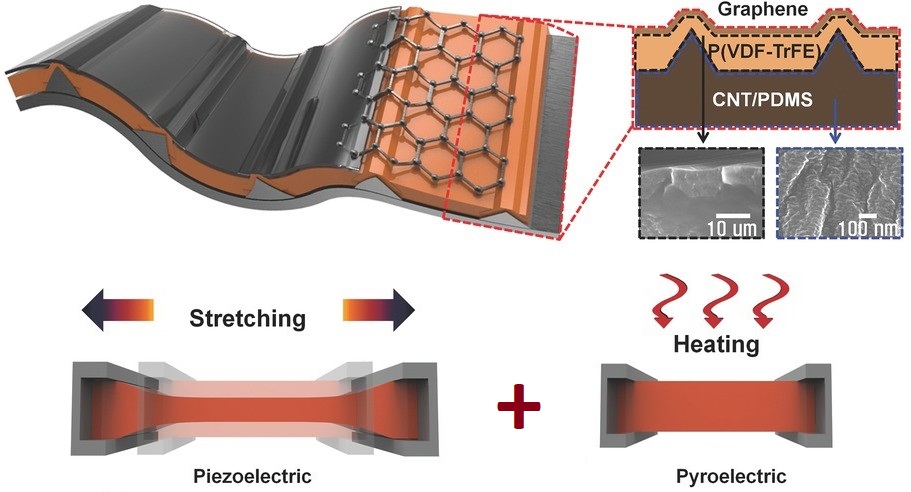}
    \caption{Illustration of a GRM-based hybrid nanogenerator, simultaneously converting kinetic and thermal energy into electricity.  Reproduced with permission from \cite{lee2014highly}.} 
    \label{fig:EH}
\end{figure}

A critical challenge in realizing IoBNT applications is continuous power provisioning \cite{kuscu2021internet}, where a self-sustaining strategy could be optimal. For micro/nanoscale devices, nanomaterial-based energy generators (or \textit{nanogenerators}) have emerged as a key element due to their ability to convert various ambient energy resources, such as motion, light, flow, and heat, into utilizable power. GRMs have been widely adopted in the development of nanogenerators (Fig.~\ref{fig:EH}) due to their excellent electrical and mechanical properties. Below, we summarize the typical energy sources harnessed by GRM-based nanogenerators for powering IoBNT applications.

\subsubsection{\textbf{EH from Motion}} In IoBNT applications, motion-based kinetic or mechanical energy can be harvested from vibrations and pressure/stress-strain variations in and around the body through piezoelectric or triboelectric nanogenerators. Yet, this process is not always straightforward. Creating electrical voltage at nanoscales using vibrations is an issue since the nano-mechanical devices rarely have a high enough resonant frequency to harvest the required energy. However, research showed that the intrinsic mechanical nonlinearity of strained nanostructured graphene could overcome this issue by acting like a double-well system when pressed \cite{lopez2011nanostructured}, thereby enabling nonlinear EH at nanoscales based on piezoelectric effect. 

GRMs have also been employed in triboelectric nanogenerators (TENGs) that capitalize on mechanical energy. \textit{Triboelectricity} is based on the principle of contact electrification between two sheets of dissimilar materials and the resulting electrostatic induction between them \cite{shi2019more}. Stretchable crumpled graphene has proven highly effective in enhancing the output performance of TENGs through the modification of surface nanostructures \cite{chen2019enhanced}. Similarly, ultra-thin GRM-based TENGs have been developed to extract human skin-generated electricity via contact with clothing or the body \cite{chu2016conformal}. Owing to their flexibility, conductivity, and transparency \cite{kim2014transparent}, GRM-based TENGs have found extensive applications in wearable electronics, hinting at their potential for use in IoBNT applications as integrated components within BNTs.

\subsubsection{\textbf{EH from Flow}} GRMs can also be used to harvest fluid-based flow energy for powering IoBNT applications. Hybrid nanogenerators that simultaneously exploit sunlight and water droplet movements have been developed using monolayer graphene films \cite{zhong2016graphene}. Other research has explored the role of the substrate in electricity generation with graphene-piezoelectric material heterostructures \cite{zhong2017graphene}. A systematic evaluation of various factors, such as graphene dimensions, materials and preparation methods, substrates, solutions, and energy outputs for ionic and non-ionic fluids, can be found in \cite{tarelho2018graphene}.

Air/wind flow can also be harnessed as an energy source for IoBNT. In \cite{wang2021boosting}, a rotating TENG based on GO-filled polyvinylidene fluoride films was developed to harvest energy from airflow. In contrast to studies where GRMs constitute the harvester or part of it, some research focuses on using graphene to enhance the performance of existing harvesters. For example, wind turbine blades coated with GO and epoxy nanocomposites demonstrated reduced lighting damages and increased energy harvesting efficiency \cite{vryonis2016improved}. Despite being implemented at macroscale, this example showcases the potential of GRMs in addressing real-world challenges, which could be adapted for micro/nanoscale IoBNT through miniaturization \cite{civas2021universal}. 

\subsubsection{\textbf{EH from Light}} GRMs have also been extensively employed in photovoltaic devices, such as solar cells, to harvest energy from sunlight or artificial light resources \cite{guo2011graphene}. In solar cells, GRMs can be used as electron and hole transport materials, buffer layers, and window or/and counter electrodes to enhance certain capabilities. One study presented a power system with series-connected organic photovoltaic cells and graphene supercapacitors on the same substrate, achieving a $\approx\!5$V open-circuit voltage suitable for flexible applications \cite{chien2015graphene}. Different from the conventional photovoltaic cells utilizing the visible light frequency range for EH, \cite{vakil2014energy} designed a graphene-based nanoscale antenna harvesting energy in the ultraviolet spectrum, showing promising results for the aerospace industry. With light being abundant both outdoors and indoors, GRM-based photovoltaic nanogenerators could support airborne IoBNT applications.

\subsubsection{\textbf{EH from Heat}} Temperature gradients in the vicinity of BNTs can also be harnessed as an energy source through thermoelectric or pyroelectric nanogenerators. \textit{Thermoelectricity} (or the Seebeck effect) is the generation of electrical energy due to a thermal difference between two dissimilar electrical conductors \cite{yu2021analysis}. One study developed a thermoelectric nanogenerator integrated with a graphene-based light-driven actuator, providing self-powered sensing capabilities \cite{zhou2021photo}. Another generator used form-stable phase change materials (PCMs) supported by rGO aerogels, with graphene nano-platelet (GNP) fillers enhancing thermal conductivity \cite{yu2021analysis}. Similarly, in \cite{mehmood2020flexible}, a flexible, conformal, and reconfigurable thin-film thermoelectric generator based on rGO-CNT composites was developed, achieving a promising thermovoltage level suitable for low-energy applications, such as bio-integrated systems. However, ensuring the bio-compatibility and durability of the employed materials is crucial, particularly for systems operating \textit{in vivo}.

\subsubsection{\textbf{EH from Sound}} Sound-based energy harvesting presents a viable option for powering IoBNT applications, with ambient acoustic vibrations and ultrasonic signals being potential energy sources. In one study, researchers leveraged GO to fabricate an acoustic nanogenerator that achieved high conversion efficiency from sound to electricity \cite{que2012flexible}. Another study introduced a stretchable, transparent nanogenerator by utilizing a polymeric piezoelectric material encapsulated between graphene electrodes, which exhibited remarkable sensitivity to various sound frequencies and amplitudes \cite{lee2013highly}. These advances indicate that GRM-based acoustic nanogenerators hold significant potential for IoBNT applications, particularly those employing liquid-borne MC in complex \textit{in vivo} and \textit{in vitro} environments, such as underwater scenarios, where conventional energy sources may be impractical or inefficient.

\subsubsection{\textbf{Hybrid EH}} While EH offers promising capabilities for BNTs, the low power density and inherent intermittency of available sources present challenges. These limitations, often due to uncontrollable factors suggest that relying on a single energy source may be insufficient for consistent operation in dynamic environments. Consequently, there is growing interest in hybrid EH approaches, which exploit multiple energy sources concurrently to boost overall energy output and enable uninterrupted functionality. 

Hybridization of energy sources is a well-established research area in the EH domain \cite{akan2017internet}, with numerous implementations at macroscale. Leveraging this accumulated knowledge, recent developments have extended hybrid EH approaches to micro/nanoscales through the use of GRM-based hybrid nanogenerators. For instance, \cite{zhong2016graphene} developed a 2D monolayer graphene film-based hybrid nanogenerator capable of harvesting energy from both sunlight and water flow simultaneously. Another study introduced a highly stretchable hybrid nanogenerator based on graphene nanosheets, which combined piezoelectric and pyroelectric properties to simultaneously harvest thermal and kinetic energy \cite{lee2014highly}. These advancements demonstrated that concurrently utilizing multiple energy sources in a complementary manner can mitigate the uncertainties associated with EH, thus paving the way for self-sustaining IoBNT applications in dynamic environments. 

However, further exploration is needed to address unique challenges and requirements arising from source hybridization, such as complexity, interoperability, and modularity. Given the existing solutions and growing research activity in this area, it is anticipated that GRM-enhanced (hybrid) nanogenerators will play a pivotal role in advancing IoBNT applications across diverse fields, including healthcare, biomedical engineering, aeronautics, automotive, aerospace, and beyond.

\begin{figure*}[t]
	\centering
	\includegraphics[width=1.7\columnwidth]{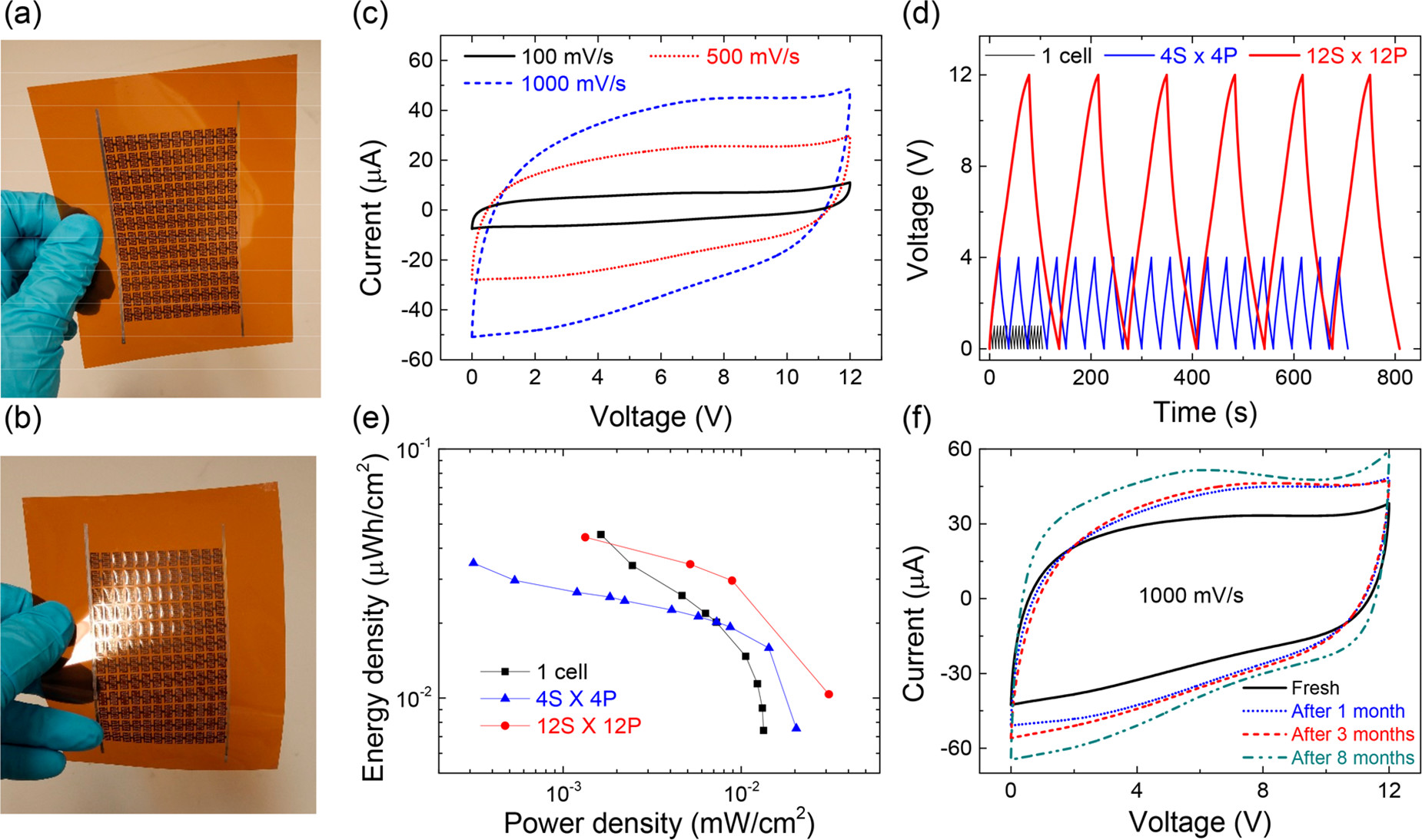}
	\caption{Large-scale integration of fully inkjet-printed flexible MSCs on Kapton: (a-b) Photographic images of MSC arrays from different perspectives. (c) Cyclic voltammetry (CV) profiles at different scan rates, captured one month post-fabrication. (f) CV profiles at various time intervals following fabrication, illustrating long-term stability.  (d) Galvanostatic charge/discharge (GCD) curves, demonstrating ideal capacitive performance. (e) Areal energy density versus power density plot, demonstrating that large-scale MSCs deliver heightened power density. Reproduced with permission
from \cite{li2017scalable}. Copyright\copyright~2017 American Chemical Society.  } 
	\label{fig:storage_scalable}
\end{figure*}

\subsection{GRM-based Micro/Nanoscale Energy Storage for IoBNT}

\begin{figure*}[t]
	\centering
	\includegraphics[width=1.6\columnwidth]{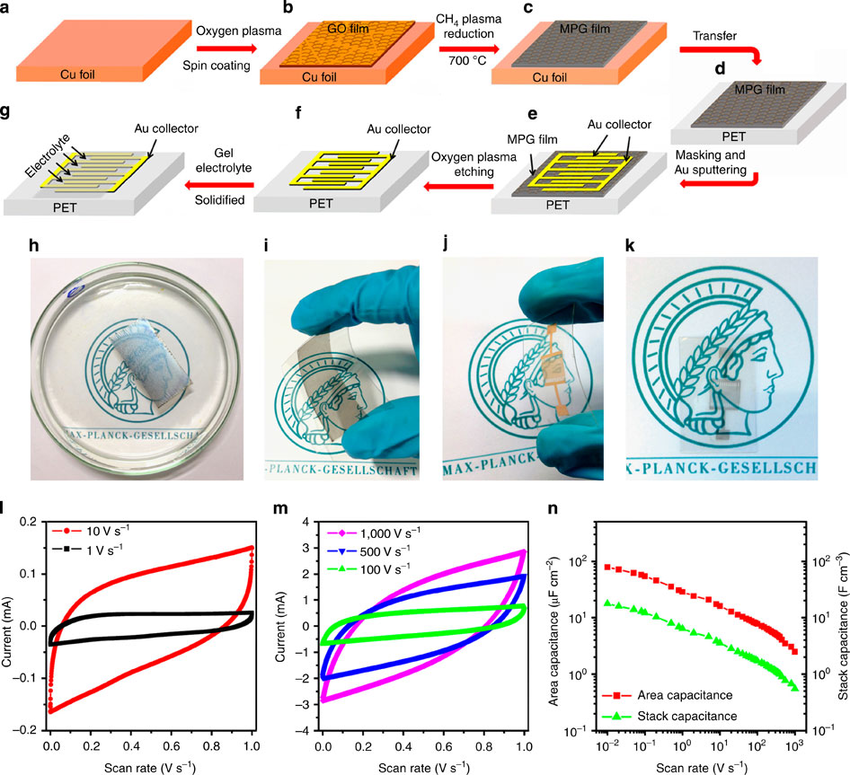}
	\caption{(a-g) Schematic illustration of the fabrication process for flexible rGO-based MSC on a PET substrate, involving plasma reduction and oxygen plasma etching. (h-k) Optical images depicting (h) the rGO film on a PMMA substrate, (i) the rGO film on a PET substrate, (j) the rGO-MSC-PET device with gold current collectors (k) the rGO-MSC-PET device without gold current collectors, illustrating the device's flexibility. (l-n) Cyclic voltammetry (CV) curves of the rGO-MSC-PET device at various scan rates, showcasing high power performance, and areal capacitance measurements, indicating high energy performance \cite{wu2013graphene_MSC}.} 
	\label{fig:storage}
\end{figure*}

In an ideal EH system, the harvested power should instantaneously match the power consumption, allowing the load or device to connect directly to the harvester without the need for intermediate components, such as storage. This concept known as \textit{power-neutral} operation \cite{sliper2020energy}, is often impractical due to the intrinsic characteristics of EH, such as uncontrollability, unpredictability, volatility \cite{beckerenergy}. As ambient energy sources are often unstable and intermittent, energy-harvesting BNTs require energy accumulation during periods of ample energy in a capacitor or battery-like storage. This stored energy is then used during periods of scarce energy to ensure continuous communications, a process referred to as \textit{energy-neutral} operation \cite{balsamo2020control}.

The growing demand for micro/nano storage components in microelectronic applications has led to the development of on-chip integrable solutions, such as the complementary micro-batteries (MBs) and micro-supercapacitors (MSCs). MBs typically exhibit high energy density, short lifetimes (typically $< 1000$ cycles \cite{jetybayeva2021recent}) and low power density. In contrast, MSCs usually possess long lifetimes (over $10^5$ cycles \cite{kyeremateng2017microsupercapacitors}), fast charge/discharge rate, high power density, but low energy density compared to MBs. However, the short lifetimes of MBs can pose significant challenges in operating regions where battery replacement or maintenance is not feasible, such as within the body. 

In addition to energy and power performance, the flexibility and compactness of energy storage components are crucial for seamless integration. Planar MCSs, such as interdigital in-plane MSCs, hold great potential for combining these attributes. For example, graphene-based planar MSCs on various substrates demonstrate remarkable scalability and integrability through various printing techniques \cite{li2017scalable, shi2019ultrahigh}, as shown in Fig.~\ref{fig:storage_scalable}. Moreover, as Wu \emph{et al.} demonstrated, graphene-based in-plane MSCs, fabricated on flexible substrates via plasma-etching, can achieve high energy densities comparable to those of lithium-ion thin-film batteries while maintaining high power densities \cite{wu2013graphene_MSC}.

MSCs typically operate based on two principles: electrostatic double-layer MSCs, which store energy through ion adsorption at the interface between the electrode material (e.g., graphene) and the electrolyte (liquid or solid-state); pseudo-MSCs, which store energy via fast reversible redox reactions between the electrode material (e.g., pseudocapacitive materials like metal oxides) and the electrolyte; and hybrid MSCs, which combine both mechanisms for rapid charge storage and high charge capacity.  

GRMs, such as graphene and rGO, are among the most widely-used active materials for planar MSC electrodes due to their high specific surface area, electrical conductivity, electrochemical stability, mechanical strength, theoretical capacitance, and ultra-thin flexible structure. Flexible graphene-based MSCs can achieve high energy and power densities, remarkable cycling stability, and mechanical robustness by leveraging these properties, as demonstrated in \cite{wu2013graphene_MSC}. In this study, GO films were reduced using methane-plasma treatment at $700^{\circ}$C, which repaired GO defects, and gold current collectors were deposited to maximize the accessible surface area of graphene, resulting in high conductivity (Fig.~\ref{fig:storage}).   

However, graphene's large specific surface area does not always translate to high capacitance, as van der Waals forces can cause graphene sheets to restack and reduce ion accessible surface area. Strategies for enhancing the energy performance of graphene-based MSCs primarily involve utilizing graphene composites and heterostructures. For instance, CNTs have been employed as nano-spacers between rGO sheets to mitigate restacking \cite{beidaghi2012micro}. Heterostructure films that combine graphene's high conductivity with the high capacity of pseudocapacitive materials have been considered to achieve high power and energy performance simulatenaously \cite{wu2017stacked}. However, optimizing the electrochemical performance of heterostructures remains challenging, as the principles governing ion and electrical transport in such architectures are not yet fully understood \cite{pomerantseva2019energy}. Moreover, it is essential to optimize the fabrication methods for graphene-based planar MSCs, as the performance of the resulting components is closely tied to the manufacturing process.

\section{Discussion on Material Challenges}
GRMs hold immense promise for facilitating the practical implementation of core IoBNT components and enhancing their performance and functionality. However, realizing this potential necessitates overcoming several material challenges inherent to GRMs, particularly in terms of reproducibility, biocompatibility, and device performance. In this section, we explore these challenges from a material science perspective, while also providing insights into potential solutions and ongoing research efforts in the field. 

IoBNT applications typically encompass a vast number of BNTs \cite{kuscu2021internet}. When implemented with GRMs, this raises concerns regarding device \textbf{\emph{heterogeneity}} in terms of performance due to existing reproducibility issues associated with GRM-based devices. The delicate and non-standardized GRM synthesis and processing methods challenge the attainment of consistent properties and functionality among the fabricated devices. Therefore, to enhance reproducibility of GRM-based BNTs and IoBNT applications, it is imperative to standardize the GRM synthesis and processing techniques. In the case of graphene, CVD is regarded as the most promising technique for producing large-area, high-quality graphene films with high electrical and thermal conductivity, as well as good structural and mechanical properties \cite{zhang2013review}. The CVD process typically involves growing graphene films on conductive metallic substrates such as copper or nickel, necessitating the transfer of graphene onto other substrates for practical device implementations. However, the graphene transfer process, particularly the wet transfer method, often subjects the graphene sample to various chemicals, resulting in the formation of defects and impurities that can stochastically alter the material properties. Hence, direct growth of graphene on the target substrate is preferable to circumvent the need for a transfer process. However, the high temperatures ($>1000 \text{ C}^{\circ}$) required for graphene synthesis via CVD can damage many types of substrates, complicating direct growth on such materials. Researchers are actively exploring low-temperature, transfer-free methods to address this challenge \cite{vishwakarma2017transfer}. 


IoBNT applications often involve integrating BNTs with complex and dynamic biological systems. Achieving this integration demands a deep understanding of the specific biological processes and tissues that interact with the devices. Additionally, it requires the development of technologies and techniques that can effectively interface with these biological systems without causing any local or systemic toxic or adverse effects. However, the \textbf{\emph{biocompatibility}} of GRMs has not yet reached a consensus, as the literature presents conflicting reports \cite{bullock2019biocompatibility}. These discrepancies regarding the biocompatibility of GRMs can be attributed to variations in material properties influenced by factors such as the specific form of graphene employed, thickness, lateral dimensions, surface chemistry, as well as the particular biological system being studied. As a result, it is crucial to explore the relationship between the physical and chemical properties of distinct graphene forms and the corresponding response elicited by the biological systems in question. A comprehensive understanding of these relationships will be instrumental in assessing the biocompatibility of GRMs and advancing their use in the development of BNTs and IoBNT applications. 

Integrating GRMs with other materials like semiconductors and insulators is essential for developing \textbf{\emph{multifunctional}} BNTs for IoBNT applications. However, this integration entails significant challenges pertaining to material compatibility, interface engineering, and fabrication processes. Discrepancies in thermal expansion coefficients, lattice constants, and chemical reactivity among different materials may induce structural defects and impair performance \cite{lemme20222d}. Achieving high-quality interfaces between different materials is crucial for efficient charge and energy transfer in BNT implementations \cite{heidari2022integrated}. Fabrication processes must be carefully designed and optimized to ensure that the integration of disparate materials does not compromise the unique properties of GRMs or other materials involved, that are core to the functionality of IoBNT.


In addition to the general challenges associated with the reproducibility and biocompatibility of GRMs, specific obstacles may emerge in the context of IoBNT components, namely micro/nano transceivers, bio-cyber interfaces, and energy harvesting/storage units. For example, GRM-based THz detectors, which are integral components of micro/nano transceivers for THz-band EM nanocommunications, commonly display low responsivity and slow operation. This is primarily due to the limited interaction between EM waves and graphene at THz frequencies, along with the inadequate absorption of EM radiation resulting from the small-scale nature of these components. To attain satisfactory responsivity, excessive bias voltages are often necessary, which is not a practical solution. Plasmonic graphene detectors can offer enhanced responsivity by augmenting light-graphene interaction via subwavelength confinement of light. However, significant potential remains for further enhancement. In this pursuit, researchers are exploring the integration of graphene with various materials and the incorporation of these composites into metamaterial structures. Despite the abundance of theoretical research in this domain, experimental demonstrations have been relatively scarce, mainly due to the reliance on sophisticated micro/nanofabrication techniques, such as patterning. Nonetheless, ongoing research efforts continue to explore and address these challenges \cite{squires2022electrically}.




While GRMs offer great potential for enabling bio-cyber interfaces in IoBNT applications thanks to their flexibility and sensitivity, certain challenges persist. These include ensuring the chemical and structural integrity of GRM-based interfaces, as well as their long-term operational stability \cite{bitounis2013prospects,san2020bioelectronics}. For both \textit{in vitro} and \textit{in vivo} IoBNT applications, maintaining the homogeneity of GRMs in terms of size, thickness, and the number of surface functional groups is crucial. Inhomogeneities in GRM structure can directly impact surface functionalization efficiency, performance evaluation, biocompatibility, and reproducibility of the resulting devices. The development of application-specific bio-functionalization methods for GRM-based bio-cyber interfaces remains a significant challenge, as it is essential to maintain the stability of these devices without altering the structure and function of biomolecules and GRMs \cite{jiang2020graphene, reina2017promises}.

Similar challenges are encountered in GRM-based EH at nanoscale. For the effective creation and mass production of nanoscale energy harvesters using GRMs, research must focus on advanced fabrication technologies that are precise, high-resolution, scalable, and cost-effective \cite{li2022graphene, pusty2022insights}. Various microfabrication techniques can be employed to fabricate microelectrodes for integration into energy harvesting BNTs. However, consistent and accurate preparation of these devices, particularly for electrode arrays, is crucial for enhancing the energy density of GRM-based nanoscale energy harvesters. Stability of the active material, adhesion between the electrode material and electrolyte, and appropriate packaging are also essential factors to consider for the development of power provisioning techniques for BNTs and IoBNT applications based on GRMs.

Specifically, TENGs, which may power IoBNT applications, face numerous challenges, including the development of efficient and scalable methods for producing large quantities of high-quality graphene \cite{cheng2022advanced, hatta2022review}. Understanding the fundamental mechanisms behind triboelectrification and charge transfer in GRMs could aid in enhancing the energy conversion efficiency of TENGs. Additionally, concerns regarding the long-term stability and performance of graphene-based TENG systems under different environmental conditions must be addressed. 

In summary, standardizing GRM synthesis and processing techniques, as well as exploring low-temperature, transfer-free methods, will play a crucial role in enhancing the reproducibility of GRM-based devices. A comprehensive understanding of the relationship between GRM properties and biological responses will be instrumental in assessing biocompatibility and furthering their use in the development of BNTs. The integration of GRMs with other materials to create multifunctional BNTs demands careful consideration of material compatibility, interface engineering, and fabrication processes. Addressing these material challenges and optimizing fabrication processes, researchers can unlock the full potential of GRMs in IoBNT applications. 

\section{Conclusions}
In this perspective, we have explored the potential of GRMs and GRM-based technologies to advance the emerging IoBNT paradigm, which particularly aims to extend human connectivity to biological environments. These environments present unique challenges, including the scale of communicating entities and the diverse forms of information exchange. Despite these challenges, we have argued that GRMs, with their remarkable properties, hold great promise for addressing these issues and enabling the development of key IoBNT components. Our discussion centered on practical transceivers of unconventional communication techniques for IoBNT applications, with a focus on GRM-based architectures capable of emitting and detecting unconventional signal forms, such as molecular and very high frequency EM signals. Examples include graphene bioFET-based MC receivers, graphene nanoporous membranes for MC transmitters, graphene plasmonic THz radiators/detectors, and GRM-based optoacoustic ultrasonic sources. Moreover, we assessed the potential of existing GRM-based bio-interfacing technologies to inform the design of sophisticated bio-cyber interfaces tailored for IoBNT applications. These technologies include biosensors, neural interfaces, and stimuli-responsive drug delivery devices. We also examined the current capabilities and limitations of GRM technologies for energy harvesting from relevant ambient sources and storing energy to provide uninterrupted power to both artificial and biological devices, i.e., BNTs, within the IoBNT ecosystem. By harnessing the exceptional properties of GRMs, the IoBNT paradigm can transform our understanding of, and interactions with biological systems, fostering unprecedented advancements in healthcare, environmental monitoring, and a myriad of other fields. 

\bibliographystyle{IEEEtran}
\bibliography{references}

\end{document}